\documentclass[letterpaper,11pt]{article}
\pdfoutput=1 % if your are submitting a pdflatex (i.e. if you have
\usepackage{jheppub} % for details on the use of the package, please
\usepackage{mathtools,enumerate}
\usepackage{multirow,array,romannum}
\usepackage[vcentermath]{youngtab}
\usepackage{tikz}
\usepackage{enumitem}
\usepackage{tikzscale}
\usepackage[all]{xy}

\def \Tr{\mbox{Tr\,}}
\def \tr{\mbox{tr\,}}
\def \ch{\mbox{ch}}

\title{\boldmath $S_N$ Orbifolds and String Interactions}

\author{Antal Jevicki,}
\author{Junggi Yoon}

\affiliation{Department of Physics, Brown University,\\Providence, RI 02912, USA}

\emailAdd{Antal\_Jevicki@brown.edu}
\emailAdd{Jung-Gi\_Yoon@brown.edu}

\preprint{{\tt BROWN-HET-1688}}

\abstract{We study interacting features of $S_N$ Orbifold CFTs. Concentrating on characters (associated with 
$S_N$ Orbifold primaries) we first formulate a novel procedure for evaluating them through $GL(\infty)_+$
tracing. The result is a polynomial formula which we show gives results equivalent to those found by Bantay. From this we deduce a hierarchy of commuting Hamiltonians featuring locality in the induced space, and nonlinear string-type interactions.}

\begin{document} 

\maketitle
\flushbottom

\section{Introduction}\label{sec:introduction}

Recently some general holographic and string theory features of 2D CFTs featuring $S_N$ orbifolds have been 
studied~\cite{Gaberdiel:2014cha,Haehl:2014yla,Belin:2014fna,Gaberdiel:2015mra,Baggio:2015jxa,Belin:2015hwa}. These are based on Large $N$ expansion of the CFT which translates into a semiclassical expansion of the dual Gravity/String theory. In particular the growth of states and phases of the partition function have been considered as pointers to a holographic string theory interpretation. For strings in 10D and 26D Minkowski space, it was demonstrated convincingly that the corresponding $S_N$ orbifold gives the correct Virasoro amplitudes~\cite{Arutyunov:1997gt,Arutyunov:1997gi}. In~\cite{Gaberdiel:2014cha,Gaberdiel:2015mra,Baggio:2015jxa}, a very interesting Higher Spin symmetry group was identified in the specific model of $\mathbb{T}^4$ Orbifolds classifying the states of tensionless strings on AdS$_3\times S^3\times \mathbb{T}^4$ through extended $\mathcal{W}$-symmetry~\cite{Henneaux:2010xg,Campoleoni:2010zq}. This greatly extends the earlier holographic studies of supersymmetric $S_N$ ($\mathbb{T}^4$/K3) orbifold CFTs representing D1-D5 branes~\cite{Maldacena:1998bw,Maldacena:1999bp,Avery:2010er,Avery:2010hs,Avery:2010vk}, where the sector of chiral primary states~\cite{deBoer:1998us} was studied in correspondence with (Super) Gravity on AdS$_3\times S^3$. Here explicit construction of all chiral primary operators and evaluation of their 3-point functions was performed in~\cite{Jevicki:1998rr,Jevicki:1998bm,Jevicki:2000it,Lunin:2000yv,Lunin:2001ew,Lunin:2001pw,Lunin:2001ne,Pakman:2007hn,Pakman:2009zz,Pakman:2009ab}. These were shown to compare with the gravitational interactions of compactified Supergravity in AdS$_3$. A concrete implementation of the `stringy exclusion principle' was also demonstrated in these constructions~\cite{Jevicki:1998rr,Jevicki:2000it}.
For the case of general orbifold CFTs recent study on the growth of states and phases of the partition function points more generally to a holographic string theory interpretation~\cite{Haehl:2014yla,Belin:2014fna,Baggio:2015jxa,Belin:2015hwa,Keller:2011xi}.

In the present paper we are concerned with general $S_N$ Orbifold CFTs with the purpose of describing interactions between general primary states. We begin by considering characters, associated with general primaries which we use as characteristic invariant variables in the $S_N$ orbifold CFT. %, their dynamics is then considered in the Large $N$ expansion. 
We formulate a novel procedure for evaluating characters in general S$_N$ orbifold CFT based on the $gl(\infty)_+$ algebra\footnote{A similar approach can be found in~\cite{Candu:2012jq} where $gl(\infty)_+$ was used to analyze the partition function of the two dimensional free boson.}. This allows one to follow some analogies with the $U(N)$ group case, and we give a representation evaluating general characters (as polynomials of a basic set). Our representation is seen to agree (in explicit comparison) with the results of Bantay~\cite{Bantay:1999us} who has given a beautiful mathematical procedure for orbifold characters through representations of the double. The procedure that we present will have some practical advantages, in particular we will use it to study the interacting features of primaries. In parallel with our earlier work~\cite{Jevicki:2013kma} involving $\mathcal{W}_N$ minimal model CFTs and the simple model of $U(N)$ we will describe the interactions in terms of a hierarchy of hamiltonians. The hamiltonians are constructed with the property that general characters appear as exact polynomial eigenfunctions. These will be demonstrated to exhibit certain locality properties in the emergent space-time. 

The content of the paper is as follows. In Section~\ref{sec:sn orbifold} we start with describing the `twisted' Hilbert space of an $S_N$ orbifold CFT representing $N$ copies of some seed (free) CFT. We define the basic set of characters associated with class primaries and give a procedure for constructing all characters as generalized Schur polynomials. To demonstrate this procedure we work out explicitely the $N$=3 case of the $S_3$ orbifold showing agreement with the expressions of Bantay. Higher cases of $N=4, 5$ are exhibited in Appendix~\ref{app:examples}.
In Section~\ref{sec:field theory of orbifold} we present the Hamiltonian(s) introduced to govern the nonlinear structure of general $S_N$ orbifold characters. We motivate this construction through the case of $U(N)$ group characters. We give the geometric interpretation of interactions through the `stringy' processes of joining and splitting. Through appropriate Fourier transforms we exhibit locality of these interactions. In Section~\ref{Application : Partition Function} we present an evaluation of partition functions, as an application of our character construction, making contact with previous works.

\newpage

\section{$S_N$ Orbifold}\label{sec:sn orbifold}

%\subsection{$S_N$ Orbifold}\label{Sec:Notations}

\subsection{Hilbert Space}\label{sec:hilbert space}

%We have seen how to generate the character of the $S_N$ orbifold in terms of Schur polynomials of $\Phi$'s in the previous section. Considering the Hilbert space of the $S_N$ orbifold, one can understand the $S_N$ orbifold character formula. 
To define Hilbert space~\cite{Dijkgraaf:1996xw} of $S_N$ orbifold CFT one starts by introducing twisted boundary conditions of primary operators\footnote{In this section, we consider only one primary operator for simplicity. In general, there can be more than one primary operator. For this case, we will later generalize the analysis in the next section. } $\phi_I$ ($I=1,2,\cdots, N$)
\begin{equation}
\phi_I(\sigma+2\pi)=\phi_{g(I)}(\sigma)\qquad (I=1,2,\cdots, N)\label{eq:twisted boundary condition1}
\end{equation}
where $g$ is an element of $S_N$, and can be represented by the partition of $N$
\begin{equation}
g=1^{\lambda_1} 2^{\lambda_2}\cdots N^{\lambda_N}\qquad \mbox{with}\quad \sum_{n=1}^N n\lambda_n=N
\end{equation}
Note that elements in the same conjugacy class of $S_N$ represent the same boundary condition. Hence, the conjugacy class $[g]$ characterizes the twisted sector $\mathcal{H}_{[g]}$ with the boundary condition $[g]$. The full Hilbert space is decomposed into the twisted sectors. In each twisted sector represented by the conjugacy class, it is sufficient\footnote{The $S_N$ invariance of the state can be realized by averaging over the $S_N$ elements. The summation over $S_N$ can be decomposed into summation over the centralizer and over the conjugacy class. At the level of character (partition function), the average over the conjugacy class is trivial.} to consider an invariant subspace under the centralizer subgroup $C_{[g]}$ of $[g]$ given by 
\begin{equation}
C_{[g]}=\prod_{n=1}^N \left( S_{\lambda_n}\times \mathbb{Z}_{n}^{\lambda_n}\right)\label{def:centralizer subgroup}
\end{equation}
For the given Hilbert space $\mathcal{H}$ of the seed CFT, the Hilbert space $\mathcal{H}_{[g]}$ of the twisted sector associated with the boundary condition $[g]$ can be written as
\begin{equation}
\mathcal{H}_{[g]}=\bigotimes_{n=1}^N\left[\left(\underbrace{\mathcal{H}_{n}\otimes \cdots \otimes \mathcal{H}_{n}}_{\lambda_n\;\text{times}}\right)^{S_{\lambda_n }}\right]\label{eq:invariant Hilbert space 1}
\end{equation}
where $\mathcal{H}_n$ is the $\mathbb{Z}_n$ invariant subspace of $\mathcal{H}^{\otimes n}$ with the boundary condition corresponding to one cycle of length $n$. Moreover, $\mathcal{M}^{S_{\lambda_n}}$ denotes the $S_{\lambda_n}$ invariant subspace of $\mathcal{M}$. 

This is true for any $S_N$ orbifold CFT whether the Hilbert space $\mathcal{H}$ of the seed CFT is a chiral sector or a full sector. Recall that we usually focus on the holomorphic part of the full Hilbert space of the two-dimensional CFTs. Thus, we will also consider the holomorphic part of the orbifold CFT. Since $S_N$ invariant states are consist of holomorphic and anti-holomorphic part, the holomorphic part is not necessarily $S_N$ invariant. Hence, one may relax the twisted boundary condition \eqref{eq:twisted boundary condition1} by introducing phase factor depending on $g$
\begin{equation}
\phi_I(z e^{2\pi})=e^{i \delta[g]}\phi_{g(I)}(z)\qquad (I=1,2,\cdots, N)\label{eq:twisted boundary condition2}
\end{equation}
as was discussed in the cyclic orbifold case~\cite{Borisov:1997nc}. The phase factor of the holomorphic part can be cancelled with that of the anti-holomorphic part. For simplicity, let us consider one cycle of length $a$ in $S_a$ for $\phi_I$ ($I=1,2,\cdots, a$) {\it i.e.}
\begin{equation}
[g]= a^1
\end{equation}
A consistent phase factor in \eqref{eq:twisted boundary condition2} is an irreducible representation $e^{{2\pi i \theta\over a}}$ of $\mathbb{Z}_a$ ($\theta=0,\cdots, a-1$). Namely, for the given cycle $[g]=a^1$, we have the following boundary condition labelled by $\theta$ $(\theta=0,1,\cdots,a-1)$.
\begin{eqnarray}
\phi_I(z e^{2\pi i})&=&e^{{2\pi i \theta\over a}}\phi_{I+1}(z)\qquad (I=1,2,\cdots, a-1)\\
\phi_a(z e^{2\pi i})&=&e^{{2\pi i \theta\over a}}\phi_{1}(z)\label{eq:twist bc ex}
\end{eqnarray}
To construct the Hilbert space satisfying the boundary condition, we define a projection operator $\mathbf{P}_{a,\theta}$ as follows.
\begin{equation}
\mathbf{P}_{a,\theta}\equiv{1\over a} \sum_{j=0}^{a-1} \exp\left[{2\pi i (-\theta +L_0)\over a}j\right]%={1\over a} \sum_{j=0}^{a-1} \exp\left[{2\pi i (-\theta +L_0-\overline{L}_0)\over a}j\right]
\end{equation}
where we rescaled the length of the space by $1/a$ ({\it e.g.} $2\pi a\; \longrightarrow \; 2\pi $). It is easy to see that $\mathbf{P}_{a,\theta}$ is indeed a projection operator\footnote{When considering the holomorphic part only, one can find a problem in \eqref{eq:projection id} for the case of a primary with non-integer conformal dimension. In this case, one may redefine $\theta$ by absorbing the conformal dimension of the primary into $\theta$. And, this prescription is related to taking ``regular part'' of characters of the seed CFT in~\cite{Bantay:1999us}.}.
\begin{equation}
\mathbf{P}_{a,\theta}^2=\mathbf{P}_{a,\theta}\quad,\quad \mathbf{P}_{a,\theta}^\dag= \mathbf{P}_{a,\theta}\quad,\quad \sum_{\theta=0}^{a-1} \mathbf{P}_{a,\theta}=1\label{eq:projection id}
\end{equation}
Hence, it projects the tensor product space $\mathcal{H}^{\otimes a}$ onto the subspace satisfying the boundary condition \eqref{eq:twist bc ex}. {\it e.g.}
\begin{equation}
e^{{2\pi i L_0\over a}}\mathbf{P}_{a,\theta}=e^{{2\pi i \theta \over a}}\mathbf{P}_{a,\theta}
\end{equation}
Using the projection operator, one can show that the trace over the subspace satisfying the twisted boundary condition \eqref{eq:twist bc ex} becomes
\begin{equation}
\tr\left(\mathbf{P}_{a,\theta} q^{{L_0\over a}}  \mathbf{P}_{a,\theta}\right)=\tr\left(q^{{L_0\over a}}  \mathbf{P}_{a,\theta}\right)={1\over a} \sum_{j=0}^{a-1} e^{ -{2\pi i \theta\over a} j} \chi\left({\tau+j \over a}\right)
\end{equation}
where $\chi(\tau)$ is the character of the seed CFT corresponding to the primary $\phi$. In general, an element $[g]$ in $S_N$ can be represented as 
\begin{equation}
 [g]=1^{\lambda_1} 2^{\lambda_2}\cdots N^{\lambda_N}\label{def:boundary condition g}
\end{equation}
In \eqref{def:boundary condition g}, each cycle of length $a$ is accompanied by the phase factor which is the irreducible representation $e^{{2\pi i \theta\over a}}$ of $Z_{a}$ ($\theta=0\cdots, a-1$). Hence, one can classify the $\lambda_a$ cycles of length $a$ according to $\theta$:
\begin{equation}
\lambda_a=\sum_{\theta=0}^{a-1} l_{a,\theta}
\end{equation}
where $l_{a,\theta}$ is the number of cycles of length $a$ with the phase factor $e^{{2\pi i \theta \over a}}$. Hence, the phase factor in \eqref{eq:twisted boundary condition2} can be represented by
\begin{equation}
\delta_{[g]}\sim\text{diag}(\underbrace{\delta_{1,1},\cdots \delta_{1,1}}_{l_{1,1}},\underbrace{\delta_{2,1},\cdots \delta_{2,1}}_{l_{2,1}},\underbrace{\delta_{2,2},\cdots \delta_{2,2}}_{l_{2,2}},\cdots, \underbrace{\delta_{N,N},\cdots \delta_{N,N}}_{l_{N,N}} )
\end{equation}
where $\delta_{a,\theta}\equiv {2\pi  \theta \over a}$. Symmetry group of the twisted sector $\mathcal{H}_{e^{i \delta[g]}g}$ is not the centralizer subgroup $C_[g]$ in \eqref{def:centralizer subgroup}, but is given by
\begin{equation}
C_{e^{i \delta[g]}[g]}=\prod_{a=1}^N\prod_{\theta=0}^{a-1} \left( S_{l_{a,\theta}}\times \mathbb{Z}_{a}^{l_{a,\theta}}\right)\label{def:new centralizer}
\end{equation}
Recall that the holomorphic part is transformed in the irreducible representation under $\mathbb{Z}_a^{l_{a,\theta}}$ action. Likewise, the holomorphic part is not necessarily invariant under $S_{l_{a,\theta}}$ in \eqref{def:new centralizer}. For instance, if both of the holomorphic and anti-holomorphic parts are fully anti-symmetric, the full state is invariant.  Hence, we will also consider irreducible representations with respect to $S_{l_{a,\theta}}$. Since the twisted sector $\mathcal{H}_{e^{i \delta[g]}g}$ can be decomposed according to the choice of $l_{a,\theta}$'s, we can write the subsector $\mathcal{H}^{\mathcal{L}}_{e^{i\delta_{[g]}}g}$ as
\begin{equation}
\mathcal{H}^{\mathcal{L}}_{e^{i\delta_{[g]}}g}=\bigotimes_{a=1}^N\left[\bigotimes_{\theta=0}^{a-1} \left(\underbrace{\mathcal{H}_{a,\theta}\otimes \cdots \otimes \mathcal{H}_{a,\theta}}_{l_{a,\theta}\;\text{times}}\right)\right]
\end{equation}
where $\mathcal{L}$ represents the specific choice of $l_{a,\theta}$ for all $a,\theta$, and will be defined fully in the next section. See \eqref{def:vec l} and \eqref{def:L}. As mentioned, note that we consider tensor products of $\mathcal{H}_{a,\theta}$ because we consider all possible irreducible representations with respect to $S_{l_{a,\theta}}$. ({\it cf}. $S_{\lambda_n}$ invariant subspace in~\eqref{eq:invariant Hilbert space 1})

The Hilbert space $\mathcal{H}_{a,\theta}$ can be viewed as an infinite-dimensional vector space, and one can define a natural $gl(\infty)_+$ action on $\mathcal{H}_{a,\theta}$. One may calculate character of $q^{{1\over a} \left(L_0-{c\over 24}\right)}$ which is an element of Cartan subgroup of the corresponding Lie group $GL(\infty)_+$. The character of $q^{{1\over a} \left(L_0-{c\over 24}\right)}$ in the fundamental representation is given by
\begin{equation}
q^{{1\over a} \left(L_0-{c\over 24}\right)}=\mbox{diag}\left(q^{\epsilon_1},q^{\epsilon_2},q^{\epsilon_3},\cdots,\right)
\end{equation}
where $\epsilon_1\leqq \epsilon_2\leqq \epsilon_3\leqq\cdots$ is the spectrum of $\mathcal{H}_{a,\theta}$. 

To calculate character of the higher representation, we consider a tensor product of $\mathcal{H}_{a,\theta}$, {\it i.e.} $(\mathcal{H}_{a,\theta})^{\otimes l_{a,\theta}}$. By Schur-Weyl duality, the tensor product $(\mathcal{H}_{a,\theta})^{\otimes l_{a,\theta}}$ can be decomposed into irreducible subspace with respects to $S_{l_{a,\theta}}$ and $GL(\infty)_+$. {\it i.e.}
\begin{equation}
(\mathcal{H}_{a,\theta})^{\otimes l_{a,\theta}}=\sum_{R_{a,\theta}\vdash l_{a,\theta}} S_{l_{a,\theta}}^{R_{a,\theta}}\otimes V_{GL(\infty)_+}^{R_{a,\theta}}
\end{equation}
where $S_{l_{a,\theta}}^{R_{a,\theta}}$ and $V_{GL(\infty)_+}^{R_{a,\theta}}$ is an irreducible subspace corresponding to the irreducible representation $R_{a,\theta}$ with respect to $S_{l_{a,\theta}}$ and $GL(\infty)_+$, repsectively. Here $R_{a,\theta}$ is a Young tableau with $l_{a,\theta}$ boxes. Hence, the twisted sector $\mathcal{H}^{\mathcal{L}}_{e^{i\delta_{[g]}}g}$ can be further decomposed according to $R_{a,\theta}$ for each $a,\theta$. We define the twisted sub-sector $\mathcal{H}^{\mathcal{L},\mathcal{R}}_{e^{i\delta_{[g]}}g}$ by
\begin{equation}
\mathcal{H}^{\mathcal{L},\mathcal{R}}_{e^{i\delta_{[g]}}g}=\bigotimes_{a=1}^N\left[\bigotimes_{\theta=0}^{a-1} \left(    S_{l_{a,\theta}}^{R_{a,\theta}}\otimes V_{GL(\infty)_+}^{R_{a,\theta}} \right)\right]\label{eq:generalized twisted sector}
\end{equation}
where $\mathcal{R}$ is a collection of $R_{a,\theta}$'s. See \eqref{def:R}.

With this preparation we can calculate characters by tracing over the irreducible subspace in the same way as is done for $U(N)$ characters. As an analogue of basic $U(N)$ traces $\phi_m= \tr(U^m)$ we introduce:
\begin{eqnarray}
 &&\sum_{v \in \mathcal{H}_{a,\theta} } \underbrace{\langle v  |\otimes \cdots\otimes \langle v|}_{m}q^{{1\over a}\left(L_0-{c\over 24}\right)} \underbrace{|v \rangle\otimes \cdots \otimes|v \rangle}_{m} = \tr_{\mathcal{H}_{a,\theta}} \left(\mathbf{P}_{a,\theta} q^{{m\over a}\left(L_0-{c\over 24}\right)}\mathbf{P}_{a,\theta} \right)\cr
&&={1\over a} \sum_{j=0}^{a-1} e^{ -{2\pi i \theta\over a} j} \chi\left({m\tau+j \over a}\right)\label{eq:analogue of phi_n}\equiv \Phi_{a,m;\theta}(\tau)
\end{eqnarray}
Then generally, as in the case of $U(N)$, the trace over the irreducible subspace $V^{R_{a,\theta}}_{GL(\infty)_+}$ leads to Schur-type  polynomials corresponding to the irreducible representation $R_{a,\theta}$ in terms of the basic set of variables with $m=1,2,\cdots$. However, note that we have to calculate the trace of $q^{L_0-{c\over 24}}$ over the irreducible subspace $S_{l_{a,\theta}}^{R_{a,\theta}}\otimes V_{GL(\infty)_+}^{R_{a,\theta}}$. Since $q^{L_0-{c\over 24}}$ acts trivially on $S_{l_{a,\theta}}^{R_{a,\theta}}$, it gives additional overall factor $\dim_R$ to the polynomial\footnote{This can be also easily obtained by using the projection operator $\mathbf{\Pi}_R$ in Section~\ref{appendix:partition function intro}}
\begin{equation}
\Tr_{S_{l_{a,\theta}}^{R_{a,\theta}}\otimes V_{GL(\infty)_+}^{R_{a,\theta}} }\left(q^{L_0-{c\over 24}}\right) = \dim_R P_{l_{a,\theta}}(R_{a,\theta})\label{eq:trace over the irreducible space S and V}
\end{equation}
where $P_n(R)$ is the Schur polynomial (of order $n$) corresponding to the representation $R$. For the invariant state in the full sector, one has to combine holomorphic part with anti holomorphic part. Note that not all combinations of them are invariant unlike fully symmetric or fully anti-symmetric cases. One has to pick up the invariant subspace from the direct product of $S^{R_{a,\theta}}_{l_{a,\theta}}$ (of the holomorphic part) and $S^{\overline{R}_{a,\theta}}_{\overline{l}_{a,\theta}}$ (of the anti-holomorphic part). It is easy to see that $R_{a,\theta}$ should be equal to $\overline{R}_{a,\theta}$ (therefore, $l_{a,\theta}=\overline{l}_{a,\theta}$) to have the invariant subspace~$ S^{0}_{l_{a,\theta}}$. Also, by using the orthogonality of the characters, one can show that the branching coefficient ${c^{R_{a,\theta}R_{a,\theta}}}_{0}$ of $ S^{0}_{l_{a,\theta}}$ is 1. {\it i.e.}
\begin{equation}
S^{R_{a,\theta}}_{l_{a,\theta}}\otimes S^{R_{a,\theta}}_{l_{a,\theta}}= S^{0}_{l_{a,\theta}}\oplus \sum_{T} {c^{R_{a,\theta}R_{a,\theta}}}_{T}S^T_{l_{a,\theta}}
\end{equation}
Therefore, the contribution of invariant states associated to the representation $R_{a,\theta}$ to the full partition function is given by
\begin{equation}
Z^{l_{a,\theta}}_{R_{a,\theta}}={1\over (\dim_R)^2} \left| \dim_R P_{l_{a,\theta}}(R_{a,\theta})\right|^2=\left|P_{l_{a,\theta}}(R_{a,\theta})\right|^2\label{eq:contribution to partition function}
\end{equation}
To establish contact with~\cite{Bantay:1999us}, we divide the (holomorphic) trace in \eqref{eq:trace over the irreducible space S and V} by $\dim_R$ in advance to define the orbifold character. Recalling the decomposition of the twisted sector in \eqref{eq:generalized twisted sector}, one can conclude that the the general character of the $S_N$ orbifold is a product of Schur polynomials of the basic set. Then, one can use the standard formula of CFT partition function where the partition function is a sum of the absolute square of characters {\it i.e.} $Z=\sum_R |\chi_R|^2$. In Section~\ref{appendix:partition function intro}, we also present another way to obtain the partition function \eqref{eq:contribution to partition function} of the twisted sector by constructing states consisting of holomorphic and anti-holomorphic parts from the beginning.

For the orbifold of chiral sector, neither the phase factor nor the arbitrary irreducible representation $R_{a,\theta}$ of $S_{l_{a,\theta}}$ is allowed because there is no anti-holomorphic part to compensate them. For this case, the only allowed collections of representations $(R_{a,\theta})$ are given by
\begin{equation}
R_{a,\theta}=\begin{cases}
1^n \quad (\; \overbrace{{\tiny\yng(2)\cdots\yng(2)}}^{n}\;) & \mbox{for} \quad \theta= 0\\
0  &\mbox{for}\quad \theta\ne 0\\
\end{cases}
\end{equation}
for $n=0,1,2,\cdots$.

\subsection{Characters of $S_N$ Orbifold}\label{sec:S_NOrbifold Characters:An Ansatz }

Denoting the (genus one) character $\chi_p\left(\tau\right)$ corresponding to a primary operator\footnote{Henceforth, primary oeperator $\phi^{(p)}$ is denoted by $p$ for simplicity.} $p$ of CFT$_2$, we define a basic set of variables $\Phi^{(p)}_{a,m;\theta}(\tau)$ :
\begin{equation}
 \Phi^{(p)}_{a,m,;\theta}(\tau)\equiv \frac{1}{a}\sum_{j=0}^{a-1}\exp\left(-{2\pi i \theta j\over  a}\right)\chi_p\left(\frac{m\tau+j}{a}\right)\quad\mbox{for}\;\begin{cases}
\;\; a=1,2,3,\cdots\\
\;\;\theta=0,1,2,\cdots,a-1\\
\;\; m=0,1,2,3,\cdots\\
 \end{cases}\label{eq:linear transformation}
\end{equation}
representing discrete Fourier transformation between the following sets:
\begin{eqnarray}
&\left\{\left. \Phi^{(p)}_{a,m;\theta }(\tau)\right|\theta =0,1,2,\cdots, a-1\right\}\qquad &(a,m\text{ : fixed})\\
&\left\{\left.\chi_p\left(\frac{m\tau+j}{a}\right)\right|j=0,1,2,\cdots, a-1\right\}\qquad &(a,m\text{ : fixed})
\end{eqnarray}
From expansion of the character $\chi_p(\tau)$
\begin{equation}
\chi_p(\tau)=q^{h_p-{c\over 24} }\sum_{j=0}^\infty d_j q^j
\end{equation}
where $h_p$ is the conformal dimension of the primary $p$ and $c$ is the central charge of the seed CFT, the basic variable $\Phi^{(p)}_{a,m;\theta }(\tau)$ takes the form
\begin{equation}
\Phi^{(p)}_{a,m;\theta }(\tau)=q^{{mh_p\over a}-{m c\over 24a}+{m\theta\over a}}\sum_{j=0}^\infty d_{aj+\theta}q^{mj}\equiv q^{m\omega(p;a,\theta) -{m c\over 24}} \sum_{j=0}^\infty d_{aj+\theta}q^{mj} \label{eq:expansion of phi}
\end{equation}
Here we define $\omega(p;a,p)$ by
\begin{equation}
\omega(p;a,\theta)\equiv \frac{1}{a}h_p+\left(a-\frac{1}{a}\right)\frac{c}{24}+\frac{\theta}{a}\label{def:omega freq}
\end{equation}

%In section~\ref{sec:Hamiltonian}, we will construct a Hamiltonian of which eigenstates are $S_N$ orbifold. For this it is natural to define $N_p(a,b)\equiv y^{(p)}_{(a,b);0}(\tau)$ which is analogous to $N=\phi_0$ in $U(N)$ case.
%
%\begin{equation}
%N_p(a,b)\equiv  y^{(p)}_{(a,b);0}(\tau)\equiv \frac{1}{a}\sum_{j=0}^{a-1}\exp\left(\frac{2\pi bi}{a}j\right)\chi_p\left(\frac{j}{a}\right)\label{def:N0}
%\end{equation}
%
%For the given `regularized' character $\chi_p\left(\tau\right)=\sum_{j=0}^\infty d_j q^j$, one can easily show that $N_p(a,b)$ is independent of $\tau$, and is the number of states in the levels which are $\left(a-b\right)$ modulo $a$. Namely,
%
%\begin{equation}
%N_p(a,b)=\sum_{j=1}^\infty d_{aj-b}
%\end{equation}
%
%Hence, it is infinite unless the number of descendants are finite\footnote{For instance, for $c=1$ limiting Virasoro minimal model, the $N_p\left(a,b\right)$ is finite.}. Recall that $N$ appears in the Hamiltonian as a coefficient of quadratic Hamiltonian for the case of $U(N)$. Since an eigenstate diagonalize both the quadratic Hamiltonian and the cubic Hamiltonian simultaneously, the eigenstate also diagonalize any linear combination of the quadratic and cubic Hamiltonians. Likewise, in the orbifold case, the character of the orbifold is still eigenstate even if we change $N_p\left(a,b\right)$ into a finite number to have a finite energy eigenvalue.

A general primary operator of $S_N$ orbifold CFT is expressed through an $N$-tuple of primaries in the seed CFT. By permuting components of the $N$-tuple, it can be written as follows.
\begin{equation}
\mathcal{P}=\langle\overbrace{p_1,p_1,\cdots,p_1}^{N_1},\overbrace{p_2,\cdots p_2}^{N_2},\overbrace{ p_3,\cdots,p_3}^{N_3}, \cdots,\overbrace{p_k,\cdots p_k}^{N_k}\rangle \qquad \mbox{with}\quad N=\sum_{j=1}^k N_j\label{eq:orbifold primary}
\end{equation}
where $p_i$ ($i=1,\cdots,k$) are distinct primaries in $\mathcal{I}$ which is a set of all primaries in the seed CFT. %For example, $S_3$ orbifold has three types of 3-tuples of primaries.
%
%\begin{equation}
%\left<p,q,r\right>\qquad \left<p,p,q\right>\qquad \left<p,p,p\right>
%\end{equation}
%
The Hilbert space associated with the primary $\mathcal{P}$ can be decomposed into subspaces for $\mathcal{P}_j\equiv\langle \underbrace{p_j,\cdots,p_j}_{N_j}\rangle$ in $S_{N_j}$ orbifold
\-\vspace{-2em}
\begin{equation}
\mathcal{H}^{S_N,\;\mathcal{P}}=\bigotimes_{j=1}^k \mathcal{H}^{S_{N_j}, \;\mathcal{P}_j}
\end{equation}
One can calculate the trace over each $\mathcal{H}^{S_{N_j}, \mathcal{P}_j}$ subspace in the same way as in Section~\ref{sec:hilbert space}. It follows that $S_N$ orbifold character denoted by $\Gamma_N(\mathcal{P})$ for the primary $\mathcal{P}$ is a product of $S_{N_i}$ orbifold characters $\Gamma_{N_i}(\mathcal{P}_j)$ for a primary $\mathcal{P}_j$. Namely,
\begin{equation}
\Gamma_N(\mathcal{P})=\prod_{i=1}^k \Gamma_{N_i}(\mathcal{P}_i)
\end{equation}
Therefore, it is sufficient to consider the character for the $N$-tuple of identical primaries, $\left<p,p,\cdots,p\right>$. Henceforth, we will consider only $S_N$ orbifold characters for identical primaries for each $N$.

A general $S_N$ orbifold character can be written as a product of the Schur polynomials of $\Phi_{a,\theta;m}(\tau)$ for each $a,\theta$ denoted by
\begin{eqnarray}
P_{n}(\tau;a,\theta;R)\equiv P_{n}\left(\left\{ \Phi_{a,1;\theta}(\tau), \Phi_{a,2;\theta}(\tau), \Phi_{a,3;\theta}(\tau),\cdots\right\};R\right)
\end{eqnarray}
where $a=1,2,\cdots$ and $\theta=0,1,2,\cdots, a-1$. $R$ is a Young tableau of $S_{n}$. It is natural to assign a degree $am$ to the variable $\Phi_{a,m;\theta}$. {\it i.e.}
\begin{equation}
\deg (\Phi_{a,m;\theta})=am 
\end{equation}
Then, the degree of the Schur polynomial of $P_{n}(a,b;R)$ is given by
\begin{equation} 
\deg\left(P_{n}(a,\theta;R)\right)=na
\end{equation}

We now introduce a simple procedure to generate general characters of $S_N$ orbifold step by step (for $\mathcal{P}=\left<p,p,\cdots,p\right>$). In~\cite{Bantay:1999us}, the $S_N$ orbifold characters were also obtained by considering a homomorphism from the fundamental group $\mathbb{Z}\oplus \mathbb{Z}$ to the permutation group $S_N$. Our procedure is based on the Hilbert space picture described in the previous subsection. It will result in a polynomial formula in terms of Schur polynomials.
\begin{enumerate}
\item Consider a partition of $N$. {\it e.g.} $N=\sum_{a=1}^N a\lambda_a$.

\item For each $\lambda_a$ ($a=1,2,\cdots, N$), find $a$-tuples of non-negative integers such that
\begin{equation}
\vec{l}_a=(l_{a,0},l_{a,1},\cdots, l_{a,a-1}) \qquad \text{with}\quad |\vec{l}_a|\equiv \sum_{\theta=0}^{a-1}l_{a,\theta}=\lambda_a\quad\text{and }\quad 0\leqq l_{a,\theta}\label{def:vec l}
\end{equation}
This corresponds to the classification of the Hilbert space according to the boundary condition (phase) in Section~\ref{sec:hilbert space}. For each $a$ and $\lambda_a$, there are the $_{\lambda_a+a-1}C_{a-1}$ number\footnote{Note that $\vec{l}_a=(l_{a,0},l_{a,1},\cdots, l_{a,a-1})$ is not a partition of $\lambda_a$. Contrast to partition of $\lambda_a$, one has to consider the order of $l_{a,\theta}$'s. For example, while there are two partitions of 2. $2=1+1=2+1$, there are 6~$\vec{l}_3$'s.
 \begin{equation}
 (2,0,0)\quad (0,2,0)\quad (0,0,2)\quad(1,1,0)\quad(1,0,1)\quad(0,1,1)
 \end{equation}} of $\vec{l}_a$'s.
For convenience, we define $\mathcal{L}$ to be $N$-tuple of $\vec{l}$'s
\begin{eqnarray}
\mathcal{L}&\equiv&\left(\vec{l}_1,\vec{l}_2,\cdots,\vec{l}_N\right)=\left(\left(l_{1,0}\right),\cdots,\left(l_{a,0},\cdots,l_{a,a-1}\right),\cdots,\left(l_{N,0},\cdots,l_{N,N-1}\right)\right)\label{def:L}
\end{eqnarray}
Note that $\left(|\vec{l}_1|,|\vec{l}_2|,\cdots,|\vec{l}_N|\right)$ is a partition of $N$ by definition.

\item For each $l_{a,\theta}$ ($\theta =0,1,2,\cdots, a-1$), choose an irreducible representation $R_{a,\theta}$ of $S_{l_{a,\theta}}$. We define $\vec{R}_a$ by
\begin{equation}
\vec{R}_a\equiv \left(R_{a,0},R_{a,1},\cdots,R_{a,a-1}\right)
\end{equation}
and
\begin{equation}
|R_{a,\theta}|\equiv(\mbox{The number of boxes in the Young tableau }R_{a,\theta})=l_{a,\theta}
\end{equation}
Furthermore, we also define $\mathcal{R}$ to be a $N$-tuple of $\vec{R}$'s :
\begin{equation}
\mathcal{R}\equiv \left(\vec{R}_a,\vec{R}_2,\cdots,\vec{R}_N\right)=\left(\left(R_{1,0}\right),\cdots,\left(R_{a,0},\cdots,R_{a,a-1}\right),\cdots,\left(R_{N,0},\cdots,R_{N,N-1}\right)\right)\label{def:R}
\end{equation}
%
%This $\mathcal{R}$ corresponds to an irreducible representation for (for $\mathcal{P}=\left<p,p,\cdots,p\right>$) of $S_N$ orbifold.

\item Finally, the $S_N$ orbifold character $\Gamma_{N}(p,\mathcal{R})$ equals
\begin{equation}
\Gamma_{N}(p,\mathcal{R})=\prod_{a=1}^N\prod_{\theta=0}^{a-1} P_{l_{a,\theta}}\left(a,\theta;R_{a,\theta}\right)\label{eq:orbifold character}
\end{equation}
\end{enumerate}
From \eqref{eq:expansion of phi}, one can see that $\Gamma_{N}(p,\mathcal{R})$ takes the form
\begin{equation}
\Gamma_{N}(p,\mathcal{R})= q^{\Delta(p,\mathcal{R})-{Nc\over 24}}\left(\rho_0+\mathcal{O}(q)\right)
\end{equation}
for some non-negative integer $\rho_0$. Here, $\Delta(p,\mathcal{R})$ is defined by (see \eqref{def:omega freq} for $\omega(p;a,\theta)$)
\begin{equation}
\Delta(p,\mathcal{R})\equiv\sum_{a=1}^N\sum_{\theta=0}^{a-1}|R_{a,\theta}|\omega(p;a,\theta)\label{def:conformal dimension}
\end{equation}
Considering the leading term of $\Gamma_{N}(p,\mathcal{R})$, one can deduce that $\Delta$ is the conformal dimension\footnote{The states corresponding to the leading term may be null~\cite{Borisov:1997nc}. {\it i.e.} $\rho_0=0$. In such a case, there is a non-negative integer shift to $\Delta$ because the (true) highest weight state in such a case comes from descendants of the seed CFT. This shift is already incorporated in the character formula \eqref{eq:orbifold character}, and is model-dependent. For example, for a given CFT character $\chi_p(\tau)=d_0+d_1 q+d_2 q^2 \cdots$, one can see that the conformal dimension of the (true) highest weight state in $\Gamma_2(p,{\tiny\yng(1,1)})$ is $2h_p+1$. {\it i.e.} 
\begin{equation}
\Gamma_2(p,({\tiny\yng(1,1)}),(0,0) )=P_2(\tau;1,1;{\tiny\yng(1,1)})=q^{2h_p-{c\over 12}}\left(d_1 q+\mathcal{O}(q^2)\right)
\end{equation}
where we assume non-degenerate vacuum ($d_0=1$) in general.
} of the primary $\mathcal{P}$.

Since $P_{l_{a,\theta}}\left(a,\theta;R_{a,\theta}\right)$ is a homogeneous polynomial of degree $a l_{a,\theta}$, the character $\Gamma_{N}(p, \mathcal{R})$ is the homogeneous polynomial of degree $N$. {\it i.e.}
\begin{equation}
\deg\left(\Gamma_{N}(p, \mathcal{R})\right)=\sum_{a=1}^{N}\sum_{\theta=0}^{a-1} a\cdot l_{a\theta}=\sum_{a=1}^N a|\vec{l}_a|=N
\end{equation}
As mentioned, a primary in $S_N$ orbifold is $N$-tuple of primaries and it can be be split into $N_j$-tuple of identical primaries ($ N_j\leqq N$ and $1\leqq j\leqq k$ for some positive integer $k$). See \eqref{eq:orbifold primary}. Following the above steps, one can choose $\mathcal{R}_j$ for each $N_j$-tuple of the identical primaries to define $\mathcal{R}_{\mathcal{P}}$ denoting the $k$-tuple of $\mathcal{R}_j$'s $(j=1,2,\cdots, k)$. {\it i.e.} $\mathcal{R}_{\mathcal{P}}=(\mathcal{R}_1,\cdots,\mathcal{R}_k)$ Then, the $S_N$ orbifold character for the primary $\mathcal{P}$ is given by
\begin{equation}
\Gamma_N(\mathcal{P},\mathcal{R}_{\mathcal{P}})=\prod_{j=1}^k \Gamma_{N_j}(\mathcal{P}_j,\mathcal{R}_j)
\end{equation}
In Section~\ref{Application : Partition Function}, we will exploit $\Gamma_N(\mathcal{P},\mathcal{R}_{\mathcal{P}})$ to calculate partition functions. We will clarify sub-sectors that we take into account for partition functions, and will show agreement with previous results~\cite{Dijkgraaf:1996xw,Bantay:2000eq,Maldacena:1999bp,Haehl:2014yla,Gaberdiel:2014cha,Belin:2014fna,Gaberdiel:2015mra,Baggio:2015jxa,Keller:2011xi}.

\subsection{Examples : $S_3$ Characters}

To demonstrate the procedure and show agreement with examples given by Bantay~\cite{Bantay:1999us} we now work out explicitly the $N=3$ case.

\begin{enumerate}
\item For $N=3$, there are three partitions of $3$.
\begin{eqnarray}
1+1+1=(3,0,0)\quad,\quad 1+2=(1,1,0)\quad,\quad 0\cdot  3=(0,0,1)
\end{eqnarray}

\item For $(3,0,0)$, we have only one choice for $\mathcal{L}$. 
\begin{equation}
\mathcal{L}=((3),(0,0),(0,0,0))
\end{equation}
The only non-zero element of $\mathcal{L}$ is $l_{1,0}=3$ so that one can choose one of 3 irreducible representations of $S_{l_{1,1}}=S_3$. Then, we have
\begin{eqnarray}
\mathcal{R}&=&(({\tiny\yng(3)}),(0,0),(0,0,0))\\
\mathcal{R}&=&(({\tiny\yng(2,1)}),(0,0),(0,0,0))\\
\mathcal{R}&=&(({\tiny\yng(1,1,1)}),(0,0),(0,0,0))
\end{eqnarray}
where $0$ represents the trivial representation. %which does not play any role in characters because $P_0=1$. 
From each $\mathcal{R}$, we can obtain the following three characters.
\begin{eqnarray}
\Gamma_3(\mathcal{R})&=&P_3(\tau;1,0;{\tiny\yng(3)})={1\over 6}\left\{\left( \Phi_{1,1;1}\right)^3+3 \Phi_{1,1;1}\Phi_{1,2;1}+2 \Phi_{1,3;1}\right\}\cr
&=&\frac{1}{6}\left\{\left(\chi\left(\tau\right)\right)^3+3\chi\left(\tau\right)\chi\left(2\tau\right)+2\chi\left(3\tau\right)\right\}
\\
\Gamma_3(\mathcal{R})&=&P_3(\tau;1,0;{\tiny\yng(2,1)})={1\over 6}\left\{2\left( \Phi_{1,1;1}\right)^3-2 \Phi_{1,3;1}\right\}\cr
&=&{1\over 3}\left\{\left(\chi\left(\tau\right)\right)^3-\chi\left(3\tau\right)\right\}\\
\Gamma_3(\mathcal{R})&=&P_3(\tau;1,0;{\tiny\yng(1,1,1)})={1\over 6}\left\{\left( \Phi_{1,1;1}\right)^3-3 \Phi_{1,1;1}\Phi_{1,2;1}+2 \Phi_{1,3;1}\right\}\cr
&=&{1\over 6}\left\{\left(\chi\left(\tau\right)\right)^3-3\chi\left(\tau\right)\chi\left(2\tau\right)+2\chi\left(3\tau\right)\right\}
\end{eqnarray}

\item For $(1,1,0)$, there are two choices for $\mathcal{L}$.
\begin{eqnarray}
\mathcal{L}=((1),(1,0),(0,0,0))\\
\mathcal{L}=((1),(0,1),(0,0,0))
\end{eqnarray}
And for each $\mathcal{L}$, there is only one choice for $\mathcal{R}$.
\begin{eqnarray}
\mathcal{R}=(({\tiny\yng(1)}),({\tiny\yng(1)},0),(0,0,0))\\
\mathcal{R}=(({\tiny\yng(1)}),(0,{\tiny\yng(1)}),(0,0,0))
\end{eqnarray}
and, they give the following characters, respectively.
\begin{eqnarray}
\Gamma_3(\mathcal{R})&=&P_1(\tau;1,0;{\tiny\yng(1)})P_1(\tau;2,0;{\tiny\yng(1)})=\Phi_{1,1;1} \Phi_{2,1;2}\cr
&=&\chi\left(\tau\right){1 \over 2}\left\{\chi\left(\frac{\tau}{2}\right)+\chi\left(\frac{\tau+1}{2}\right)\right\}\\
\Gamma_3(\mathcal{R})&=&P_1(\tau;1,0;{\tiny\yng(1)})P_1(\tau;2,1;{\tiny\yng(1)})= \Phi_{1,1;1} \Phi_{2,1;1}\cr
&=&\chi\left(\tau\right) {1\over 2}\left\{\chi\left(\frac{\tau}{2}\right)-\chi\left(\frac{\tau+1}{2}\right)\right\}
\end{eqnarray}

\item  For $(0,0,1)$, we have three choices for $\mathcal{L}$, and the corresponding possible $\mathcal{R}$ is given by
\begin{eqnarray}
\mathcal{L}=((0),(0,0),(1,0,0))\quad\Longrightarrow\quad\mathcal{R}=((0),(0,0),({\tiny\yng(1)},0,0))\\
\mathcal{L}=((0),(0,0),(0,1,0))\quad\Longrightarrow\quad \mathcal{R}=((0),(0,0),(0,{\tiny\yng(1)},0))\\
\mathcal{L}=((0),(0,0),(0,0,1))\quad\Longrightarrow\quad \mathcal{R}=((0),(0,0),(0,0,{\tiny\yng(1)}))
\end{eqnarray}
Hence, a character corresponding to each $\mathcal{R}$ equals
\begin{eqnarray}
\Gamma_3(\mathcal{R})&=&P_1(\tau;3,0;{\tiny\yng(1)})=\Phi_{3,1;0}\cr
&=&{1\over 3}\left\{\chi\left(\frac{\tau}{3}\right)+\chi\left(\frac{\tau+1}{3}\right)+\chi\left(\frac{\tau+2}{3}\right)\right\}\\
\Gamma_3(\mathcal{R})&=& P_1(\tau;3,1;{\tiny\yng(1)})= \Phi_{3,1;1}\cr
&=&{1 \over 3}\left\{\chi\left(\frac{\tau}{3}\right)+e^{-\frac{2\pi i}{3}}\chi\left(\frac{\tau+1}{3}\right)+e^{\frac{2\pi i}{3}}\chi\left(\frac{\tau+2}{3}\right)\right\}\\
\Gamma_3(\mathcal{R})&=&P_1(\tau;3,2;{\tiny\yng(1)})=\Phi_{3,1;2}\cr
&=&{1\over 3}\left\{\chi\left(\frac{\tau}{3}\right)+e^{\frac{2\pi i}{3}}\chi\left(\frac{\tau+1}{3}\right)+e^{-\frac{2\pi i}{3}}\chi\left(\frac{\tau+2}{3}\right)\right\}
\end{eqnarray}

\end{enumerate}
These agree with the $S_3$ orbifold characters found in~\cite{Bantay:1999us}. We present further examples of $S_4$ and $S_5$ orbifold characters in Appendix~\ref{app:examples}.

\section{Field Theory of the $S_N$ Orbifold}\label{sec:field theory of orbifold}

Our expression in Section~\ref{sec:sn orbifold} gives a general $S_N$ orbifold character which is a product of the Schur polynomials of $\Phi_{a,m;\theta}$'s. Based on this representation we can now deduce an effective Hamiltonian for the characters. Considering Fock space of $\Phi_{a,m;\theta}$'s, we will establish a Hamiltonian which can be diagonalized by the orbifold characters. The motivation for this representation comes from an analogous construction known for the characters of $U(N)$ which we first summarize as the `toy model'.

\subsection{Toy Model : $U\left(N\right)$ Character}\label{sec:toy model}

We start with the Casimir $C_2$ of $U\left(N\right)$ given by the expression:
\begin{equation}
C_2\left(R\right)={1\over 2}(R, R+2\rho)={1\over 2} \sum_{j=1}^N  r_j\left( r_j+N-(2j-1)\right)
\end{equation}
where $R=\sum_i r_i \epsilon_i$ is an irreducible representation of $U(N)$ in the orthogonal basis $\left\{\epsilon_i|i=1,2,\cdots,N\right\}$, and $\rho$ is the Weyl vector defined to be $\sum_i\left(\frac{N+1}{2}-i\right)\epsilon_i$. The well known Weyl character formula is given by
\begin{equation}
\Theta\left(R\right)=\frac{\sum_{\omega\in W}\epsilon\left(\omega\right)e^{\omega\left(\lambda+\rho\right)}}{\sum_{\omega\in W}\epsilon\left(\omega\right)e^{\omega\rho}}
\end{equation}
The Hamiltonian $H$ which we are introducing is a representation of the Casimir operator so that the characters are (Schur polynomial) eigenstates:
\begin{equation}
H\Theta\left(R\right)=C_2\left(R\right)\Theta\left(R\right)
\end{equation}

For this one uses the basic set of variables $\phi_n$ given by
\begin{equation}
\phi_n\equiv\sum_{k=1}^n\left(-1\right)^{k+1}\Theta(R_k^{\left(n\right)})
\end{equation}
where $R_k^{\left(n\right)}$ is a Young tableau defined by $r_i$ which is the number of boxes in the $i$th row as follows.
\begin{equation}
r_i=\begin{cases}
n-k+1&,i=1\\
1&,i=2,\cdots, k\\
0&,i=k+1,\cdots,n\\
\end{cases}
\qquad \left(k=1,2,\cdots,n\right)
\end{equation}
For example,
\begin{eqnarray}
\phi_3&=&\Theta\left(\;{\tiny\yng(3)}\;\right)-\Theta\left(\;{\tiny\yng(2,1)}\;\right)+\Theta\left(\;{\tiny\yng(1,1,1)}\;\right)\\
\phi_4&=&\Theta\left(\;{\tiny\yng(4)}\;\right)-\Theta\left(\;{\tiny\yng(3,1)}\;\right)+\Theta\left(\;{\tiny\yng(2,1,1)}\;\right)-\Theta\left(\;{\tiny\yng(1,1,1,1)}\;\right)
\end{eqnarray}
One can easily check that $\phi_n$ is a power-sum symmetric polynomial of order $n$. Th $U(N)$ character of the irreducible representation $R$ is a Schur polynomial of $\phi_n$'s
\begin{equation}
\Theta\left(R\right)=P_{|R|}(R;\{\phi_i\})=\frac{1}{n!}\sum_{g\in S_{|B|}}\left[\ch_R(g)\prod_{i=1}^\infty\left(\phi_i\right)^{\lambda(g)_i}\right]
\end{equation}
where $\ch_R(g)$ is the character of $g$ in the irreducible representation $R$ of the permutation group $S_{|R|}$. Decomposing the Casimir into a leading (and subleading) terms (in 1/$N$) 
\begin{equation}
C_2\left(R\right)={N\over 2}\left[\sum_{j=1}^N  r_j+\frac{1}{N}\sum_{j=1}^N  r_j\left( r_j-(2j-1)\right)\right]
\end{equation}
one can define unperturbed quadratic hamiltonian $H_2$ and can represent the perturbation as $H_3$ so that: 
\begin{equation}
\left(H_{2}+H_{3}\right)\Theta\left(R\right)=C_2\left(R\right)\Theta\left(R\right)
\end{equation}
Clearly the quadratic Hamiltonian is easily found to be
\begin{equation}
H_2\equiv {N\over 2}\sum_{n=1}^\infty n\phi_n\frac{\partial}{\partial \phi_n}
\end{equation}
giving the desired unperturbed energy
\begin{equation}
H_2\Theta\left(R\right)={N|R|\over 2}\Theta\left(R\right)
\end{equation}
where $|R|$ is the total number of boxes in the Young tableau $R$. {\it i.e.} $|R|\equiv\sum_{i=1}^N r_i$

There are several different ways to obtain the subleading 1/$N$ generating operator $H_3$. From the (fermionic) Weyl representation of $U(N)$ characters it can be deduced through bosonization~\cite{Jevicki:1991yi}. Following a group theoretic way, one can use the fusion rules of conjugacy classes of the $S_N$ group to derive the same form~\cite{Nomura:1986}. One has
\begin{equation}
H_3=\sum_{n=2}^{\infty}\sum_{m=1}^{n-1}{n \over 2} \phi_m\phi_{n-m}\frac{\partial}{\partial \phi_n}+\sum_{n=1}^{\infty}\sum_{m=1}^{\infty}{nm\over 2} \phi_{n+m}\frac{\partial}{\partial \phi_n}\frac{\partial}{\partial \phi_m}
\end{equation}
where the eigenvalue corresponding to $\Theta(R)$ is
\begin{equation}
H_3\Theta\left(R\right)=\left({1\over 2}\sum_{j=1}^N r_j( r_j-2j+1)\right)\Theta\left(R\right)\label{eq:cubic interaction u(n)}
\end{equation}
It is instructive to see the geometrical interpretation of the cubic interaction. For this we interpret the variable $\phi_n$ (power-sum symmetric polynomial) as a (closed) loop with winding number $n$. Then, the first term in the cubic interaction \eqref{eq:cubic interaction u(n)} is splitting of one loop into two loops. The second term represents joining of two loops into one loop. Moreover, the coefficients of the cubic interaction is natural because they correspond to the number of all possible ways in such a action (splitting or joining, respectively).

In sum, we have the total Hamiltonian:
\begin{equation}
H\equiv H_2+H_3={N\over 2}\sum_{n=1}^\infty n\phi_n\frac{\partial}{\partial \phi_n}+\sum_{n=2}^{\infty}\sum_{m=1}^{n-1}{n\over 2}\phi_m\phi_{n-m}\frac{\partial}{\partial \phi_n}+\sum_{n=1}^{\infty}\sum_{m=1}^{\infty}{nm\over 2}\phi_{n+m}\frac{\partial}{\partial \phi_n}\frac{\partial}{\partial \phi_m}
\end{equation}
and, its eigenvalue is the Casimir of the corresponding Young tableau. 
\begin{equation}
H\Theta\left(R\right)=C_2\left(R\right)\Theta\left(R\right)
\end{equation}
%
%This Hamiltonian was also derived by the collective field theory of $U(N)$ matrix model~\cite{Jevicki:1991yi} in which the matrix model Hamiltonian was expressed in terms of invariants $\phi_n$'s. 
Above we have motivated the construction of the Hamiltonian starting from the fact that the characters are explicitly (Schur) polynomial eigenstates. For the $S_N$ orbifold, we have expressed, in the previous section, the characters as generalized Schur polynomials. From this we can infer the form of the Hamiltonian operator(s).

\subsection{Hamiltonians for $S_N$ Orbifold Characters}\label{sec:Hamiltonian}

In the case of $U(N)$ one had a sequence of commuting Hamiltonians $H_n$ representing an integrable hierarchy which are essentially given as $\tr(P^m)$ in the matrix model description. We will now describe a similar Hamiltonian hierarchy for the $S_N$ orbifold theory. 

We consider a Fock space consisting of all possible $\Phi_{a,m;\theta}$'s as in the $U(N)$ case. %, or equivalently, of all $S_N$ orbifold characters for $N=0,1,2,\cdots$.
In this Fock space, one may define Hamiltonian hierarchies $H^{\text{orb}}_n$ which are diagonalized by the $S_N$ orbifold characters $\Gamma_N(\mathcal{P},\mathcal{R})$ with some eigenvalue $E^{\text{orb}}_n(\mathcal{P},\mathcal{R})$ :
\begin{equation}
H^{\text{orb}}_n\Gamma_N(\mathcal{P},\mathcal{R})=E^{\text{orb}}_n(\mathcal{P},\mathcal{R})\Gamma_N(\mathcal{P},\mathcal{R})
\end{equation}
and, they commute each other
\begin{equation}
[H^{\text{orb}}_n,H^{\text{orb}}_m]=0
\end{equation}
so that the orbifold character simultaneously diagonalize them. By generalizing the Hamiltonians in the $U(N)$ example, we now give the Hamiltonian hierarchies, for $n=2,3$ describing the $S_N$ orbifold.
\begin{eqnarray}
H^{\text{orb}}_2 &\equiv&\sum_{p\in\mathcal{I}}\sum_{a=1}^\infty\sum_{\theta=0}^{a-1}\sum_{m=1}^\infty m\omega(p;a,\theta)\Phi^{(p)}_{a,m;\theta} \frac{\partial}{\partial  \Phi^{(p)}_{a,m;\theta} }\\
H^{\text{orb}}_3 &\equiv& \sum_{p\in\mathcal{I}}\sum_{a=1}^\infty\sum_{\theta=0}^{a-1}\epsilon_{p;a,\theta}\left[\sum_{n=2}^\infty\sum_{m=0}^{n-1} {n\over 2} \Phi^{(p)}_{a,m;\theta} \Phi^{(p)}_{a,n-m;\theta }\frac{\partial}{\partial  \Phi^{(p)}_{a,n;\theta }}\right.\cr
&&\qquad\qquad\qquad\qquad\left.+  \sum_{n,m=1}^\infty {nm\over 2} \Phi^{(p)}_{a,n+m;\theta } {\partial^2\over \partial  \Phi^{(p)}_{a,n;b } \partial  \Phi^{(p)}_{a,m;\theta }}\right]\label{def:cubic interaction of orbifold}
\end{eqnarray}
where $\epsilon_{p;a,\theta}$ is an arbitrary constant. Moreover, the eigenvalues corresponding to $\Gamma_N(\mathcal{P},\mathcal{R})$ are given by
\begin{eqnarray}
E^{\text{orb}}_2(\mathcal{P},\mathcal{R})&=&\Delta(\mathcal{P},\mathcal{R})\equiv \sum_{p\in\mathcal{P}} \Delta(p,\mathcal{R})\\
E^{\text{orb}}_3(\mathcal{P},\mathcal{R})&=&\sum_{p\in\mathcal{I}}\sum_{a=1}^N\sum_{\theta=0}^{a-1}\sum_j{\epsilon_{p;a,\theta}\over 2} (R_{a\theta})_j\left[(R_{a\theta})_{j}-2j+1\right]
\end{eqnarray}
Analogous expressions can be given for higher Hamiltonians. The proof for the above follows from two observations :
\begin{enumerate}[label=\textit{\roman*})]
\item The orbifold character is a product of the Schur polynomials, and each Schur polynomial $P_{|R_{a,\theta}|}(p;a,\theta;R_{a,\theta})$ is a function of $\Phi^{(p)}_{a,m;\theta}$ for $m=1,2,\cdots $. ($p,a,\theta$ : fixed)

\item The Schur polynomial simultaneously diagonalizes the quadratic and cubic Hamiltonians of the $U(N)$ case.
\end{enumerate}
We emphasize that there is an ambiguity in choosing the coefficients $\omega(p;a,\theta)$ and $\epsilon_{p;a,\theta}$. Recall that that the eigenvalue of the quadratic Hamiltonian is proportional to the number of boxes in Young tableau. Among the physical quantities in the permutation orbifold, it is the conformal dimension $\Delta(p,\mathcal{R})$ in \eqref{def:conformal dimension} that is proportional to the number of the Young tableaux $\mathcal{R}$. Therefore, we chose $\omega(p;a,\theta)$ in \eqref{def:omega freq} as the coefficient of the quadratic Hamiltonian. On the other hand, unlike the case of $U(N)$ character, physical meaning of the eigenvalue $E^{\text{orb}}_3(\mathcal{P},\mathcal{R})$ is not clear. Thus, we leave the constant $\epsilon_{p;a,\theta}$ undetermined for now.

%Note that each term of the quadratic Hamiltonian $H^{\text{orbifold}}_2$ is proportional to the infinite number $N_p(a,b)$. Recall that, in the $U(N)$ model, the quadratic Hamiltonian $H_2$ commutes with the cubic Hamiltonian $H_3$. Since a $U(N)$ character diagonalize both $H_2$ and $H_3$, it is also an eigenstate of $aH_2+b H_3$ for arbitrary constants $a$ and $b$. Likewise, one can change $N_p(a,b)$ into some finite number to define `renormalized' Hamiltonian which is still diagonalized by orbifold characters. Then, eigenvalues of the `renormalized' Hamiltonian become finite. One of the natural choices for such a finite number is $\omega(p;a,b)$. Then, the `renormalized' quadratic Hamiltonian is given by
%

%
%and the eigenvalue of $H^{\text{orb}}_2$ corresponding to the character $\Gamma_N(\mathcal{P},\mathcal{R})$ is the `naive' conformal dimension $\Delta(\mathcal{P},\mathcal{R})$. As in the $U(N)$ case, the quadratic Hamiltonian $H^{\text{orb},ren}_2$ has degeneracy, which agrees with the degenerate `naive' conformal dimension found in \cite{Bantay:1999us}. The most degeneracies are split by the cubic interaction $H^{\text{orb}}_3$. By renormalizing $N_p(a,b)$, the eigenvalue of the `renormalized' full Hamiltonian $H^{\text{orb}}_2+H^{\text{orb}}_3$ reads
%
%\begin{equation}
%E^{\text{orb}}(\mathcal{P},\mathcal{R})=\sum_{a=1}^N\sum_{b=1}^a E_{R_{ab}}\left(R_{ab};\omega(p;a,b)\right)
%\end{equation}

\subsection{Geometrical Interpretation}\label{sec:Geometrical Interpretation}

In Section~\ref{sec:toy model}, we have seen the geometrical meaning of $U(N)$ character. That is, $\phi_m$ represents a loop with winding number $m$ and the cubic Hamiltonian splits and joins these loops. The Schur polynomial of $\phi_m$'s representing these loops is an eigenstate of both of quadratic and cubic Hamiltonians.

\begin{figure}[t!]
  \centering
  \[
   \begin{minipage}[h]{0.20\linewidth}
        \vspace{0pt}
        \includegraphics[width=\linewidth]{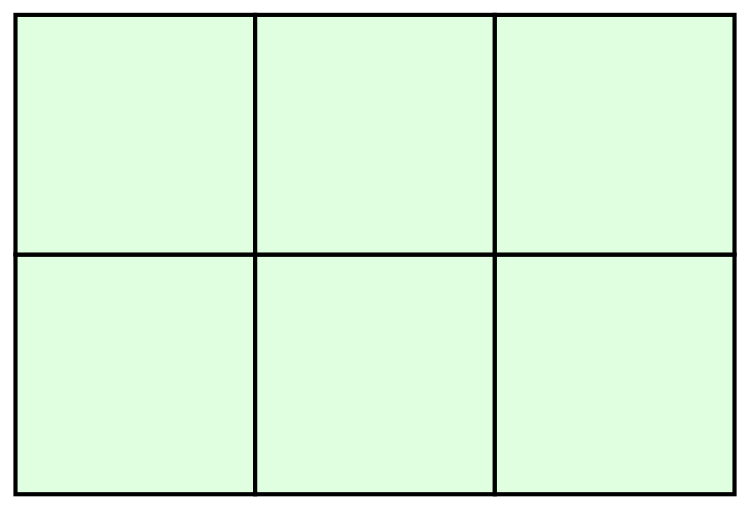}
        { \centering $\chi_p\left({2\tau\over 3}\right)$ \par}
   \end{minipage} + e^{-{2\pi i \theta\over 3}}   \begin{minipage}[h]{0.265\linewidth}
        \vspace{0pt}
        \includegraphics[width=\linewidth]{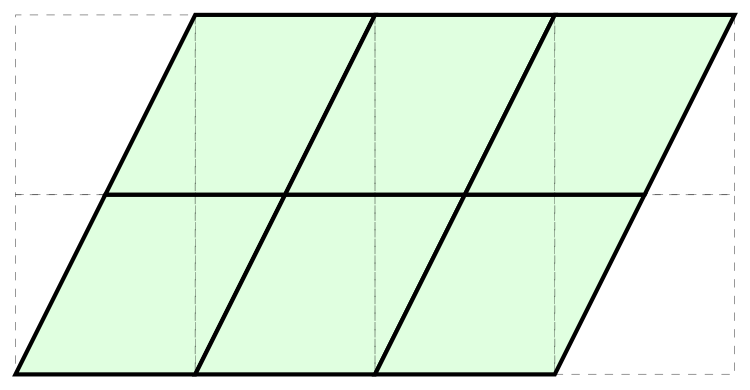}
        { \centering $\chi_p\left({2\tau+1\over 3}\right)$ \par}
   \end{minipage}+e^{-{4\pi i \theta\over 3}}\begin{minipage}[h]{0.335\linewidth}
        \vspace{0pt}
        \includegraphics[width=\linewidth]{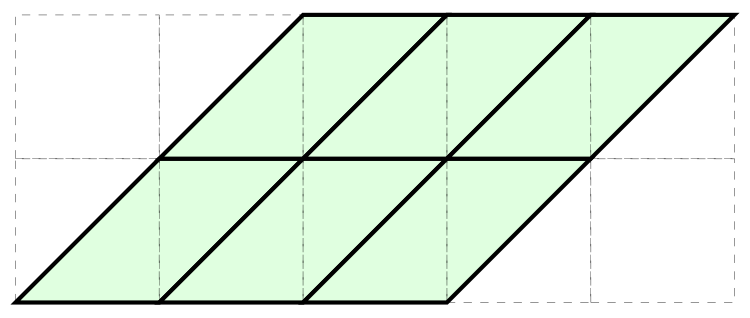}
        { \centering $\chi_p\left({2\tau+2\over 3}\right)$ \par}
   \end{minipage}
   \]\\[-2em]

   \[
 =  \begin{minipage}[h]{0.20\linewidth}
        \vspace{0pt}
        \includegraphics[width=\linewidth]{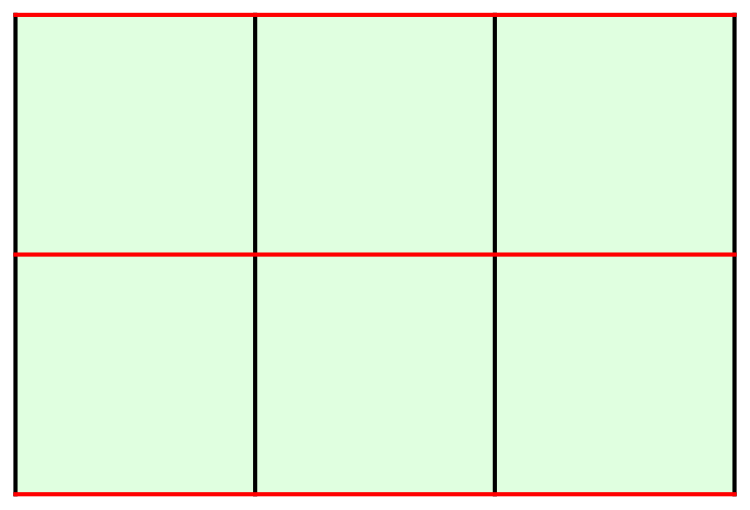}
        { \centering $\Phi^{(p)}_{3,2;\theta}$ \par}
   \end{minipage} \longrightarrow  \begin{minipage}[h]{0.40\linewidth}
        \vspace{0pt}
        \includegraphics[width=\linewidth]{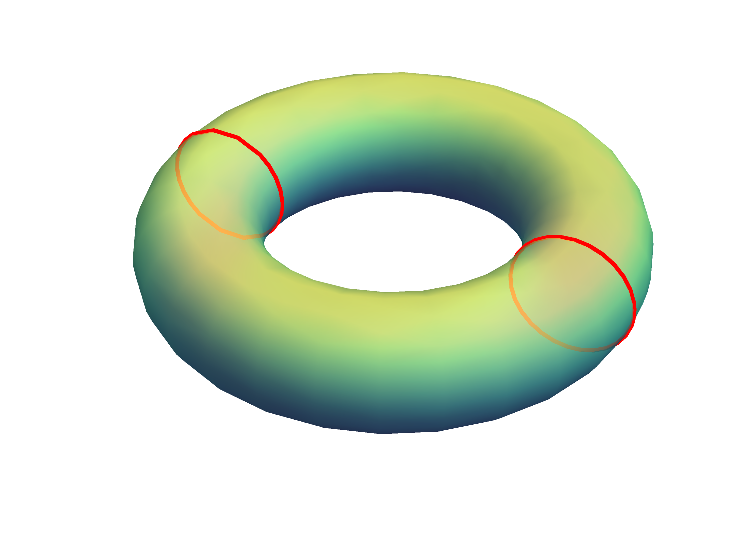}
   \end{minipage}
  \]
  \caption{Pictorial Interpretation of $\chi_p\left({m\tau+j\over a}\right)$ and $\Phi^{(p)}_{a,m;\theta}$. This represents the discrete Fourier transformation between $\chi_p\left({2\tau+j\over 3}\right)$ and $\Phi^{(p)}_{3,2;\theta}$ in \eqref{eq:linear transformation}. The cubic Hamiltonian $H_3^{\text{orb}}$ joins and split tori along the horizontal line (red line).}\label{fig:torus}
\end{figure}

Since the $S_N$ orbifold character is also a product of Schur polynomials of $\Phi^{(p)}_{a,m;\theta}$, one may put a similar geometrical interpretation on $\Phi^{(p)}_{a,m;\theta }$. Namely, one may think $\Phi^{(p)}_{a,m;\theta}$ as a loop with winding number $m$ like $\phi_m$. However, we need to make an interpretation on other labels such as $a$ and $\theta$. For this we visualize $\chi_p\left({m\tau +j\over a}\right)$ as a torus associated a primary $p$ where there are two ``winding numbers''. As mentioned, we already assigned $m$ to one of the winding numbers. We will assign $a$ to the other winding number of the torus which we will call ``dual winding number''. In addition, $j$ is associated with how the torus is twisted. Then, \eqref{eq:linear transformation} implies that $\Phi^{(p)}_{a,m;\theta}$ is (discrete) Fourier transformed torus with respect to ``twist number''~$j$. $\theta$,~which is conjugate to twist number $j$, is the analogue of the crystal momentum in the lattice. One way to show such a torus $\Phi^{(p)}_{a,m;\theta}$ graphically is to draw a box diagram with $m$ rows and $a$ columns where the left and the right sides are glued. At the same time, the top and the bottom sides are twisted according to $\theta$, and then are sewn together. See figure~\ref{fig:torus}. For convenience, let us denote such a box diagram by $(p; a,m;\theta)$. This geometrical interpretation is consistent with the illustrations shown in~\cite{Haehl:2014yla}. From \eqref{eq:linear transformation}, one can see that the box diagrams $(p;a,m,b)$ for $\theta=0,1,2,\cdots,a-1$ are conjugate to $(a\times m)$ box diagrams twisted by $j=0,1,\cdots a-1$. See figure~\ref{fig:torus}. 
From the Parseval's theorem of the (discrete) Fourier transformation, one has 
\begin{equation}
\sum_{\theta=0}^{a-1} \left|  \mbox{box diagram}\; (p;a,m;b)\right|^2=\sum_{j=0}^{a-1} \left| (a\times m)\;\mbox{box diagram twisted by }j\right|^2
\end{equation}
After summing over all possible primaries $p\in \mathcal{I}$, we obtain one of the identities that are useful in calculation of partition functions.
\begin{equation}
\sum_{p\in\mathcal{I}}\sum_{\theta=0}^{a-1}\left|\Phi^{(p)}_{a,m;\theta }\right|^2={1\over a}\sum_{j=0}^{a-1} Z_{CFT}\left({m\tau +j \over a}\right)
\end{equation}

\begin{figure}[t!]
  \centering
  \[
  \begin{tabular}{ c m{1.5em} c}
  \multirow{2}{*}{ \begin{minipage}[h]{0.2\linewidth}
        \vspace{20pt}
        \includegraphics[width=\linewidth]{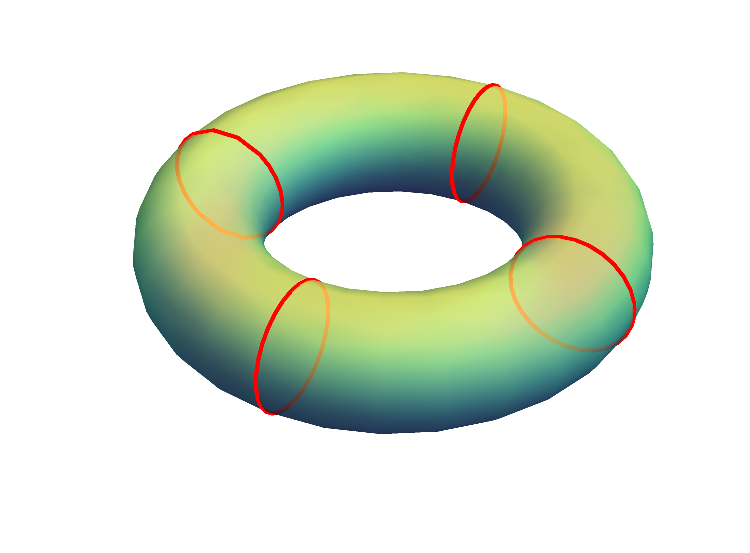}
        \centering $\Phi^{(p)}_{3,4;\theta}$
   \end{minipage}  } &\begin{tikzpicture}   \draw[thick,->] (0,0) -- (0.75,0.3);\end{tikzpicture}& $\;\;4\times$\begin{minipage}[h]{0.2\linewidth}
        \vspace{0pt}
        { \centering $4\Phi^{(p)}_{3,3;\theta}\Phi^{(p)}_{3,1;\theta}$ \par}
        \includegraphics[width=\linewidth]{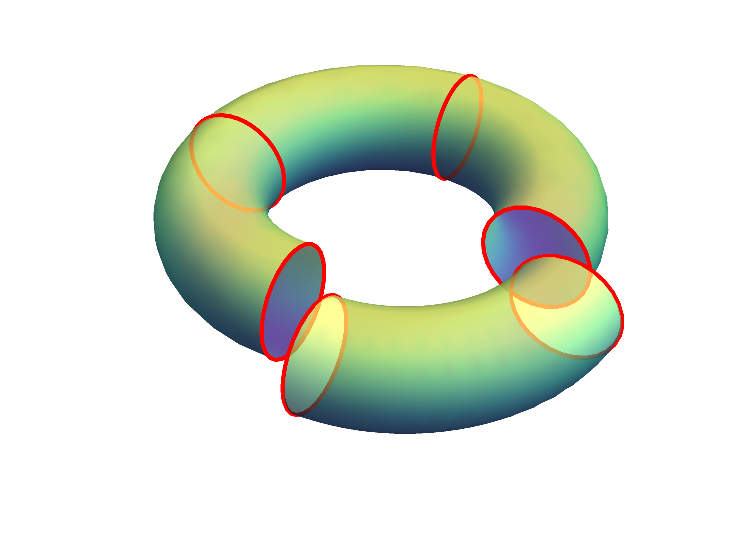}
   \end{minipage}$\begin{tikzpicture}   \draw[thick,->] (0,0) -- (0.75,0);\end{tikzpicture}$
 $\;\;4\times $\begin{minipage}[h]{0.30\linewidth}
        \vspace{0pt}
        { \centering $4\Phi^{(p)}_{3,3;\theta}\Phi^{(p)}_{3,1;\theta}$ \par}
        \includegraphics[width=\linewidth]{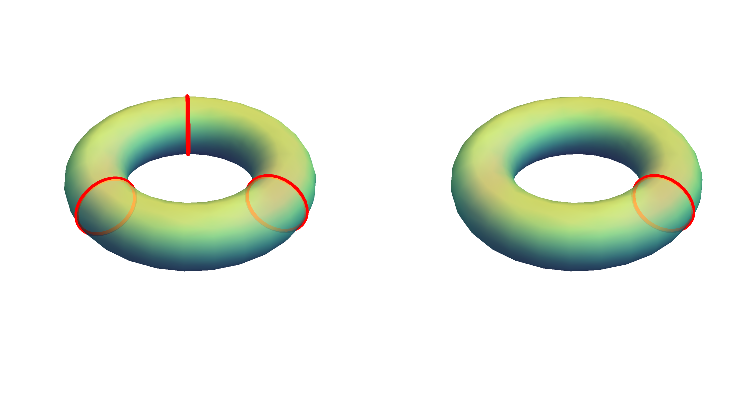}
   \end{minipage}\\
 &\begin{tikzpicture}  \draw[thick,->] (0,0) -- (0.75,-0.3);\end{tikzpicture}   &  $\;\; 2\times$\begin{minipage}[h]{0.2\linewidth}
        \vspace{0pt}
        \includegraphics[width=\linewidth]{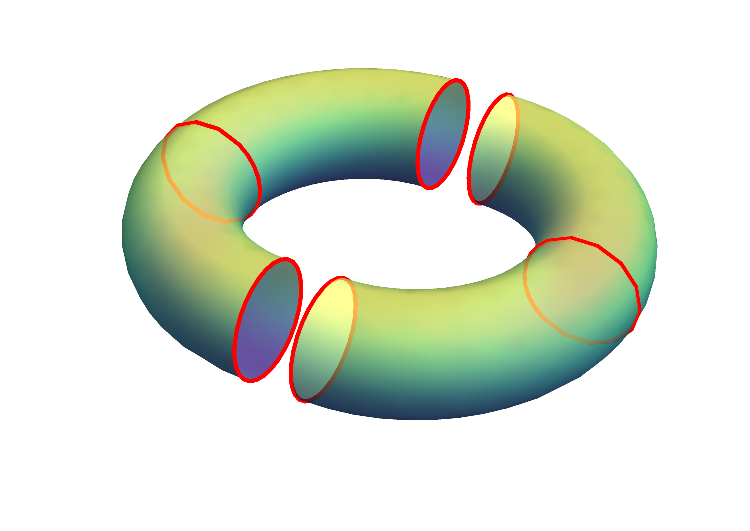}
        { \centering $2\Phi^{(p)}_{3,2;\theta}\Phi^{(p)}_{3,2;\theta}$ \par}
   \end{minipage}$\begin{tikzpicture}   \draw[thick,->] (0,0) -- (0.75,0);\end{tikzpicture}$
 $\;\;2\times $\begin{minipage}[h]{0.30\linewidth}
        \vspace{0pt}
        \includegraphics[width=\linewidth]{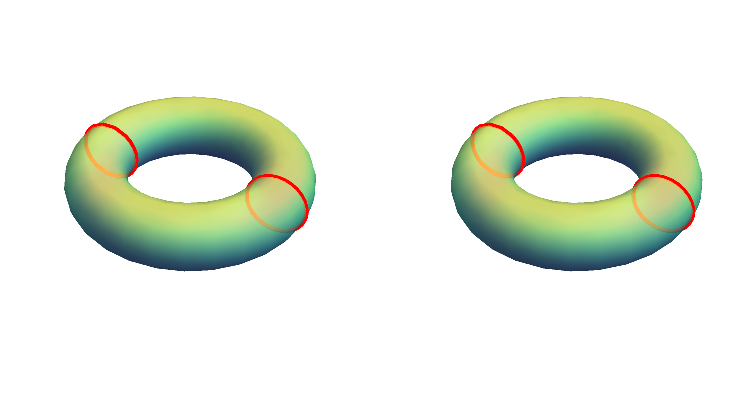}
        { \centering $2\Phi^{(p)}_{3,2;\theta}\Phi^{(p)}_{3,2;\theta}$ \par}
   \end{minipage}\\
   \end{tabular}
  \]
  \caption{Splitting of the torus. The coefficients correspond to the number of all possible ways to cut torus along two red lines.}\label{fig:splitting}
\end{figure}
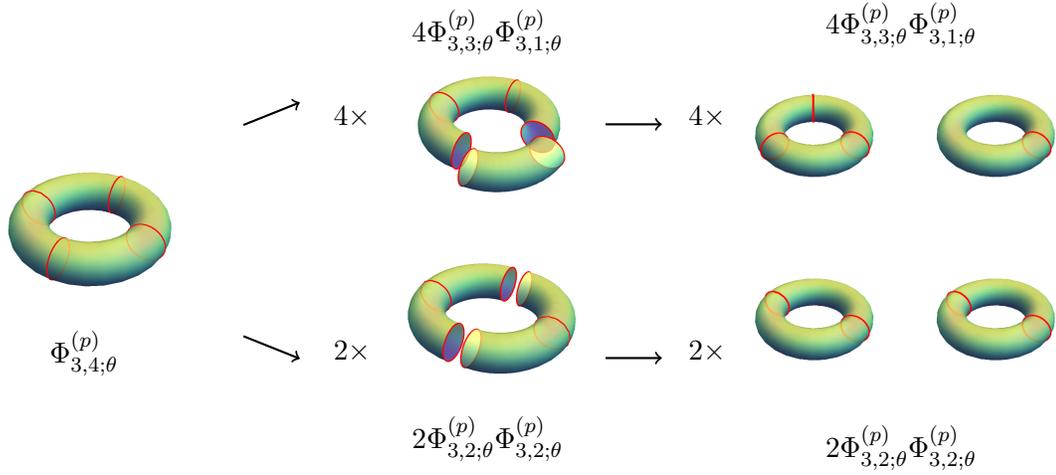

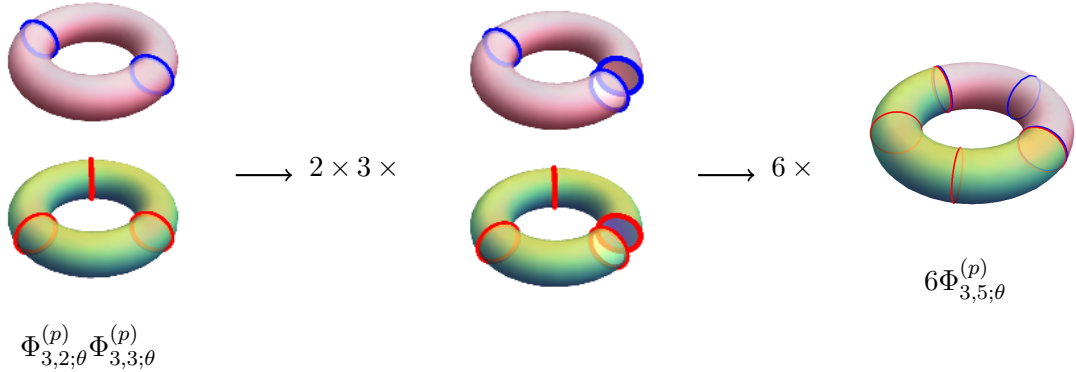
\begin{figure}[t!]
  \centering
  \[
  \begin{minipage}[h]{0.25\linewidth}
        \vspace{0pt}
        \includegraphics[width=\linewidth]{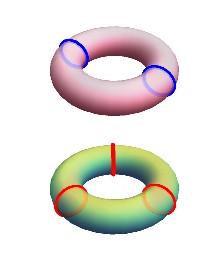}
        { \centering $\Phi^{(p)}_{3,2;\theta}\Phi^{(p)}_{3,3;\theta}$ \par}
   \end{minipage}\begin{tikzpicture}   \draw[thick,->] (0,0) -- (0.75,0);\end{tikzpicture} \;\; 2\times 3\times\begin{minipage}[h]{0.25\linewidth}
        \vspace{0pt}
        \includegraphics[width=\linewidth]{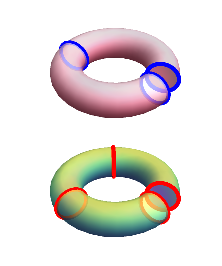}
        { \centering $\;$ \par}
   \end{minipage}\begin{tikzpicture}   \draw[thick,->] (0,0) -- (0.75,0);\end{tikzpicture} \;\; 6\times \begin{minipage}[h]{0.25\linewidth}
        \vspace{0pt}
        \includegraphics[width=\linewidth]{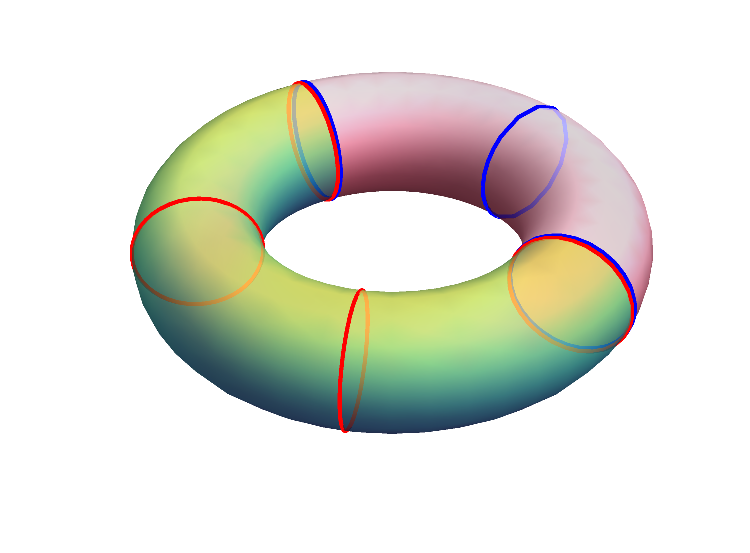}
        { \centering $6\Phi^{(p)}_{3,5;\theta}$ \par}
   \end{minipage}
  \]
  \caption{Joining of tori. The coefficients correspond to the number of all possible ways to cut each torus along one red lines.}\label{fig:joining}
\end{figure}

Similar to $U(N)$ case, the cubic Hamiltonian $H_3^{\text{orb}}$ of the orbifold also splits and joins each torus $\Phi^{(p)}_{a,m;\theta}$ with respect to the winding number $m$ (along the red line in the figure~\ref{fig:torus}). Graphically, the first term of $H_3^{\text{orb}}$ in \eqref{def:cubic interaction of orbifold} represents splitting of a torus $(p;a,m;\theta)$ into two tori $(p;a,m-n;\theta)$ and $(p;a,n;\theta)$. See figure~\ref{fig:splitting}. In addition, the second term represents joining of two tori $(p;a,m;\theta)$ and $(p;a,n;\theta)$ into a torus $(p;a,n+m;\theta)$. See figure~\ref{fig:joining}. Note that the Hamiltonians of orbifold does not change the ``dual winding number'' $a$, the ``conjugate twist number'' $\theta$ or primary $p$ in $\Phi^{(p)}_{a,m;\theta}$. Hence, tori with different $a,\theta$ and $p$ are decoupled. 

\subsection{Locality}

In the case of D1-D5 brane system captured by the symmetric product orbifold of $\mathbb{T}^4$ or $K3$, the chiral primary operator was studied~\cite{Jevicki:1998rr,Jevicki:1998bm,Lunin:2000yv,Lunin:2001ew,Lunin:2001pw,Lunin:2001ne,Pakman:2007hn}. In this case, a local interaction were shown to appear and were constructed in~\cite{Jevicki:2000it}.

The interaction features locality in the emerging coordinates. Recall the $\theta$ was interpreted as the Fourier transformation of $j$ in \eqref{eq:linear transformation}. The interaction is local in $\theta$. The other quantum number $m$ features splitting and joining processes which are characteristic of matrix model interactions. By Fourier transformation $e^{im\varphi}$, one can see the locality in $\varphi$. All together we have $\Phi^{(p)}(a,\varphi,\theta)$ with locality in the three coordinates $(a,\varphi,\theta)$.
\begin{equation}
H^{\text{orb}}_3\sim \sum_{p\in \mathcal{I}}\sum_{a=1}^\infty \sum_{\theta=0}^{a-1}\epsilon_{p;a,\theta} \int d\varphi \left[\Phi^{(p)}(a,\varphi,\theta)\right]^3
\end{equation}

\section{Application : Partition Functions}\label{Application : Partition Function}

In this section, we apply the construction in Section~\ref{sec:sn orbifold} to the evaluation and study of partition functions. First, we consider a construction $S_n$ invariant states containing both holomorphic and anti-holomorpic parts. Based on these we can calculate the partition functions, and show that the result agrees with the one that follows from the use of characters ({\it e.g.} See~\eqref{eq:contribution to partition function}). We also consider partition functions in special cases to make contact with earlier studies.

\subsection{Partition Functions}\label{appendix:partition function intro}

For the given holomorphic and anti-holomorphic Hilbert space ($\mathcal{H}_{a,\theta}$ and $\overline{\mathcal{H}}_{a,\theta}$), one may consider them as infinite vector spaces, and define a natural $gl(\infty)_+$ Lie algebra acting on each vector space. In the corresponding group $GL(\infty)_+$, $q^{L_0-{c\over 24}}$ and $\bar{q}^{\overline{L}_0-{c\over 24}}$ are elements of the Cartan subgroup.
\begin{eqnarray}
q^{L_0-{c\over 24}}&=&\mbox{diag}\left(q^{\epsilon_1},q^{\epsilon_2},q^{\epsilon_3},\cdots\right)\qquad \left(\;\epsilon_1\leqq \epsilon_2\leqq \epsilon_3\leqq \cdots\right)\\
\bar{q}^{\overline{L}_0-{c\over 24}}&=&\mbox{diag}\left(\bar{q}^{\bar{\epsilon}_1},\bar{q}^{\bar{\epsilon}_2},\bar{q}^{\bar{\epsilon}_3},\cdots\right)\qquad \left(\;\bar{\epsilon}_1\leqq \bar{\epsilon}_2\leqq \bar{\epsilon}_3\leqq \cdots\right)
\end{eqnarray}
where $\epsilon_j$ and $\overline{\epsilon}_k$ is the spectrum of the holomorphic and anti-holomorphic sector, respectively. We now consider the holomorphic and anti-holomorphic parts together. A state in the (full) Hilbert space is denoted by
\begin{equation}
|m,\overline{m}\rangle\in \mathcal{H}_{a,\theta}\otimes\overline{\mathcal{H}}_{a,\theta}
\end{equation}
where $m\in\{\epsilon_1,\epsilon_2,\cdots\}$ and $\overline{m}\in\{\bar{\epsilon}_1,\bar{\epsilon}_2,\cdots\}$. The trace of $q^{L_0-{c\over 24}} \bar{q}^{\overline{L}_0-{c\over 24}}$ over the Hilbert space $\mathcal{H}_{a,\theta}\otimes\overline{\mathcal{H}}_{a,\theta}$ becomes
\begin{equation}
\sum_{m,\overline{m}} \langle m,\overline{m} | q^{L_0-{c\over 24}} \bar{q}^{\overline{L}_0-{c\over 24}} |m,\overline{m} \rangle=\Phi_{a,\theta;1}\overline{\Phi}_{a,\theta;1}
\end{equation}
For the tensor product of the Hilbert spaces, $\left(\mathcal{H}_{a,\theta}\otimes\overline{\mathcal{H}}_{a,\theta}\right)^{\otimes n}$, one can define $S_n\times S_n$ action by
\begin{equation}
(\sigma,\overline{\sigma})|\{m_j\},\{\overline{m}_j\}\rangle\equiv |m_{\sigma(1)},\overline{m}_{\overline{\sigma}(1)}\rangle\otimes\cdots \otimes |m_{\sigma(n)},\overline{m}_{\overline{\sigma}(n)}\rangle
\end{equation}
where $|\{m_j\},\{\overline{m}_j\}\rangle$ is a general state in $\left(\mathcal{H}_{a,\theta}\otimes\overline{\mathcal{H}}_{a,\theta}\right)^{\otimes n}$ defined by
\begin{equation}
|\{m_j\},\{\overline{m}_j\}\rangle\equiv |m_1,\overline{m}_1\rangle\otimes\cdots \otimes |m_n,\overline{m}_n\rangle
\end{equation}
It is convenient to define three commuting projection operators $\mathbf{\Pi}_R, \overline{\mathbf{\Pi}}_R$ and $\mathbf{\Upsilon}_R$ which project Hilbert space onto irreducible representation $R$.
\begin{equation}
\mathbf{\Pi}_R\equiv {1\over n!}\sum_{\sigma\in S_n}\dim_R \ch_R(\sigma)(\sigma,1)\quad,\quad \overline{\mathbf{\Pi}}_R\equiv {1\over n!}\sum_{\overline{\sigma}\in S_n}\dim_R \ch_R(\overline{\sigma})(1,\overline{\sigma})
\end{equation}
and
\begin{equation}
\mathbf{\Upsilon}_R \equiv{1\over n!}\sum_{\sigma\in S_n}\dim_R \ch_R(\sigma)(\sigma,\sigma)
\end{equation}
They satisfy the properties of projection operator. {\it e.g.}
\begin{equation}
\mathbf{\Pi}_R \mathbf{\Pi}_T=\delta_{R,T} \mathbf{\Pi}_R\quad,\quad \mathbf{\Pi}_R^\dag=\mathbf{\Pi}_R\quad,\quad \sum_{R\vdash n} \mathbf{\Pi}_R=1\label{eq:projection properties}
\end{equation}

First of all, in order to calculate partition function of the $S_n$ invariant subspace, we define $S_n$ invariant state
\begin{eqnarray}
|\{m_j\},\{\overline{m}_j\}\rangle_{\text{inv}}\equiv\mathbf{\Upsilon}_{0}  |\{m_j\},\{\overline{m}_j\}\rangle\label{eq:full invariant state}
\end{eqnarray}
The trace over such states gives the desired result (See \eqref{eq:sum of Schur polynomial})
\begin{eqnarray}
Z^{n}&=&\sum_{\{m_j\},\{\overline{m}_j\}} \langle \{m_j\},\{\overline{m}_j\}| q^{L_0-{c\over 24}} \bar{q}^{\overline{L}_0-{c\over 24}}|\{m_j\},\{\overline{m}_j\}\rangle_{\text{inv}}\cr
%&=&{1\over n!}\sum_{\rho\in S_n}\sum_{\{m_j\},\{\overline{m}_j\}} \langle \{m_j\},\{\overline{m}_j\}| q^{L_0-{c\over 24}} \bar{q}^{\overline{L}_0-{c\over 24}}(\rho,\rho)|\{m_j\},\{\overline{m}_j\}\rangle\cr
&=&{1\over n!}\sum_{\rho\in S_n}   \prod_{k=1}^n \Phi_{a,\theta;k}^{\rho_k} \overline{\Phi}_{a,\theta;k}^{\rho_k} =\sum_{\{\rho_k\}\vdash n}\prod_{k=1}^n {\Phi_{a,\theta;k}^{\rho_k}  \overline{\Phi}_{a,\theta;k}^{\rho_k} \over \rho_k! k^{\rho_k}}\label{eq:full partition function n}
\end{eqnarray}
where for the given $\rho\in S_n$, $\{\rho_k\}$ is defined by $\rho=1^{\rho_1} 2^{\rho_2}\cdots n^{\rho_n}$. 

As in Section~\ref{sec:hilbert space}, one may consider irreducible subspace with respect to $S_n$. By using the projection operator $\overline{\mathbf{\Pi}}_R$, we project the invariant states in \eqref{eq:full invariant state} onto the irreducible representation $R$ with respect to the anti-holomorphic part.
\begin{equation}
|R;\{m_j\},\{\overline{m}_j\}\rangle\equiv \mathbf{\Upsilon}_{0}\overline{\mathbf{\Pi}}_R |\{m_j\},\{\overline{m}_j\}\rangle=\sum_{\rho,\overline{\sigma}\in S_n}{ \dim_R\over (n!)^2} \ch_R(\overline{\sigma})(\rho,\rho\overline{\sigma})|\{m_j\},\{\overline{m}_j\}\rangle\label{eq:anti-holomorphic inv state}
\end{equation}
Note that this state is also $S_n$ invariant because $\mathbf{\Upsilon}_{0}$ and $\overline{\mathbf{\Pi}}_R$ commute with each other. Using the third identity in \eqref{eq:projection properties}, one can further decompose the state \eqref{eq:anti-holomorphic inv state} into irreducible representation of $S_n$ with respect to the holomorphic part.
\begin{eqnarray}
|R;\{m_j\},\{\overline{m}_j\}\rangle&=&{1\over (n!)^3}\sum_{T\vdash n}\sum_{\rho,\sigma,\overline{\sigma}\in S_n} \dim_R \dim_T \ch_T(\sigma) \ch_R(\overline{\sigma})(\sigma\rho,\overline{\sigma}\rho)|\{m_j\},\{\overline{m}_j\}\rangle\cr
%&=&{1\over (n!)^3}\sum_{T\vdash n}\sum_{\rho,\sigma,\overline{\sigma}\in S_n} \dim_R \dim_T \ch_T(\sigma\overline{\sigma}^{-1}\rho^{-1}) \ch_R(\rho)(\sigma,\overline{\sigma})|\{m_j\},\{\overline{m}_j\}\rangle\cr
&=&{1\over (n!)^2}\sum_{\sigma,\overline{\sigma}\in S_n} \dim_R \ch_R(\sigma \overline{\sigma}^{-1}) (\sigma,\overline{\sigma})|\{m_j\},\{\overline{m}_j\}\rangle
\end{eqnarray}
where we used the orthogonality of the characters. This implies that the non-zero contribution only comes from the irreducible representation $R$ of the holomorphic part. Hence, one may write $|R;\{m_j\},\{\overline{m}_j\}\rangle$ as
\begin{equation}
|R;\{m_j\},\{\overline{m}_j\}\rangle=\mathbf{\Upsilon}_{0}\mathbf{\Pi}_R \overline{\mathbf{\Pi}}_R |\{m_j\},\{\overline{m}_j\}\rangle
\end{equation}
The trace over this irreducible subspace is given by
\begin{eqnarray}
Z^n_R&=&\sum_{\{m_j\},\{\overline{m}_j\}} \langle R;\{m_j\},\{\overline{m}_j\}| q^{L_0-{c\over 24}}\bar{q}^{\overline{L}_0-{c\over 24}}|R;\{m_j\},\{\overline{m}_j\}\rangle\cr
%&=&\sum_{\{m_j\},\{\overline{m}_j\}} \langle \{m_j\},\{\overline{m}_j\}| q^{L_0-{c\over 24}}\bar{q}^{\overline{L}_0-{c\over 24}} \mathbf{\Upsilon}_{0}\mathbf{\Pi}_R \overline{\mathbf{\Pi}}_R |\{m_j\},\{\overline{m}_j\}\rangle\cr
%&=&{1\over (n!)^2} \sum_{\sigma,\overline{\sigma}\in S_n} \dim_R \ch_R(\sigma\overline{\sigma}^{-1}) \langle \{m_j\},\{\overline{m}_j\}| q^{L_0-{c\over 24}}\bar{q}^{\overline{L}_0-{c\over 24}} (\sigma,\overline{\sigma}) |\{m_j\},\{\overline{m}_j\}\rangle\cr
&=&{1\over (n!)^2} \sum_{\sigma , \overline{\sigma} \in S_n} \dim_R \ch_R(\sigma\overline{\sigma}^{-1}) \prod_{k=1}^n \Phi_{a,\theta;k}^{\sigma_k} \overline{\Phi}_{a,\theta;k}^{\overline{\sigma}_k}\label{eq:partition function of n, R}
\end{eqnarray}
where $\sigma=1^{\sigma_1} \cdots n^{\sigma_n}$ and $\overline{\sigma}=1^{\overline{\sigma}_1} \cdots n^{\overline{\sigma}_n}$. One can easily show that the summation of $ Z^n_R$ over the irreducible representation $R$ gives the full partition function in \eqref{eq:full partition function n}.
\begin{equation}
Z^n=\sum_R Z^n_R=\sum_{\{\sigma_k\}\vdash n}\prod_{k=1}^n {\Phi_{a,\theta;k}^{\sigma_k} \overline{\Phi}_{a,\theta;k}^{\sigma_k}\over \sigma_k! k^{\sigma_k}}
\end{equation}
Note that $\{\sigma_k\}$ and $\{\overline{\sigma}_k\}$ in \eqref{eq:partition function of n, R}, which are the partitions of $n$, depend only on the conjugacy class. Hence, one can average the coefficient $\ch_R(\sigma\overline{\sigma}^{-1})$ in \eqref{eq:partition function of n, R} over the conjugacy class. Then, one can show that
\begin{eqnarray}
&&{\dim_R\over \left|{\scriptstyle[\sigma]}\right|\left|{\scriptstyle[\overline{\sigma}]}\right|}\sum_{\rho\in[\sigma]}\sum_{\overline{\rho}\in [\overline{\sigma}]}\ch_R(\rho\overline{\rho}^{-1})={\dim_R\over n!} \sum_{g\in S_n} ch_R (g^{-1}\sigma g \overline{\sigma}^{-1})\cr
&=&{\dim_R\over n!} \sum_{g\in S_n} D^R_{ij}(g^{-1}) D^R_{jk}(\sigma) D^R_{kl}(g)D^R_{li}(\overline{\sigma}^{-1})=\ch_R(\sigma) ch_R(\overline{\sigma}^{-1})
\end{eqnarray}
Therefore, one can conclude that
\begin{equation}
Z^n_R= {1\over (n!)^2} \sum_{\sigma,\overline{\sigma}\in S_n} \ch_R(\sigma) ch_R(\sigma)  \prod_{k=1}^n \Phi_{a,\theta;k}^{\sigma_k} \overline{\Phi}_{a,\theta;k}^{\overline{\sigma}_k}=|P_n(a,\theta;R)|^2
\end{equation}
which is the same as the result \eqref{eq:contribution to partition function} from the $S_N$ orbifold character.

\subsection{Examples}

In this section, we will calcuate $S_N$ orbifold partition function from the orbifold character $\Gamma_N(\mathcal{P},\mathcal{R})$. One can express the partition function $Z^N_{\text{orbifold}}$ of the $S_N$ orbifold in terms of $\Gamma_N(\mathcal{P},\mathcal{R})$ as follows.
\begin{equation}
Z^N_{\text{orbifold}}=\sum_{\mathcal{P}}\sum_{\mathcal{R}_{\mathcal{P}}} \left|\Gamma_N(\mathcal{P},\mathcal{R}_{\mathcal{P}})\right|^2
\end{equation}
where the summations run over all possible $N$-tuple of primaries $\mathcal{P}$ and over the all possible Young tableaux $\mathcal{R}_{\mathcal{P}}$. One can define a generating function for $Z^N_{\text{orbifold}}$ :
\begin{equation}
Z_{\text{orbifold}}=\sum_{N=0}^\infty t^N Z^N_{\text{orbifold}}
\end{equation}
where $Z^0_{\text{orbifold}}=1$ for consistency. Using an identity of the Schur polynomial
 \begin{equation}
 \sum_{\substack{R\\|R|=n}}\left|P_{n}(R;\{x_j\})\right|^2=\sum_{\{\lambda_j\}\vdash n} \quad \prod_{j=1}^n\left[{1 \over \lambda_j!} \left({1\over j}\left|x_j\right|^2\right)^{\lambda_j}\right]\label{eq:sum of Schur polynomial}
 \end{equation}
it is not difficult to prove that
\begin{eqnarray}
Z_{\text{orbifold}}&=&\prod_{p\in\mathcal{I}} \left[\sum_{N=0}^\infty \sum_{\mathcal{R}}\prod_{a=1}^N\prod_{\theta=0}^{a-1} t^{a|R_{a,\theta}|}\left|P_{|R_{a,\theta}|}(p;a,\theta,R_{a,\theta})\right|^2\right]\cr
&=&\exp\left[\sum_{M=1}^\infty t^M T_M Z_{CFT}(\tau,\overline{\tau})\right]
\end{eqnarray}
where $ Z_{CFT}(\tau,\overline{\tau})$ is the partition function of the seed CFT and $T_M $ is $M$th-Hecke operator of which action is defined as follows. 
\begin{equation}
T_M Z_{CFT}(\tau,\overline{\tau})\equiv {1\over M}\sum_{a \;|\; M}\sum_{j=0}^{a-1} Z_{CFT}({M\tau +ja\over a^2},{M\overline{\tau} +ja\over a^2})
\end{equation}
This agrees with~\cite{Bantay:2000eq,Keller:2011xi,Haehl:2014yla,Belin:2014fna}.

One can also consider a partition function of a sub-sector of the orbifold CFT. The simplest sub-sector is the untwisted sector where all components of representation\footnote{A representation $\mathcal{R}_{\mathcal{P}}$ for a primary $\mathcal{P}$ (see \eqref{eq:orbifold primary}) is a $k$-tuple of representations $\mathcal{R}_j=((R_j)_{1,0},((R_j)_{2,0},(R_j)_{2,1}),\cdots,)$ for $N_j$-tuple of identical primaries of $S_{N_j}$ orbifold $(1\leqq j\leqq k)$. } $\mathcal{R}_j$ in $\mathcal{R}_{\mathcal{P}}$ are 0 except for $(R_j)_{1,0}$ for all $j=1,\cdots k$. Therefore, the partition function of untwisted sector is given by
\begin{equation}
Z^N_{\text{untwisted}}=\sum_{\mathcal{P}}\sum_{\mathcal{R}_{\mathcal{P}}} \left|\Gamma_N(\mathcal{P},\mathcal{R}_{\mathcal{P}})\right|^2\qquad \mbox{where}\quad (\mathcal{R}_{j})_{a,\theta}=\begin{cases}
R & \mbox{for}\quad (a,\theta)= (1,0)\\
0 &\mbox{for}\quad (a,\theta)\neq(1,0)\\
\end{cases}\label{eq:untwist partition function}
\end{equation}
where the summation runs over the all possible Young tableaux $R$. We also construct a generating function for the untwisted partition function.
\begin{equation}
Z_{\text{untwisted}}=\sum_{N=0}^\infty t^N Z^N_{\text{untwisted}}
\end{equation}
Using the orbifold characters, one can show that
\begin{equation}
Z_{\text{untwisted}}=\prod_{p} \left[\sum_{N=0}^\infty \sum_{\substack{R\\|R|=N}}  t^{|R|}\left|P_{|R|}(p;1,0,R)\right|^2\right]=\exp\left[\sum_{m=1}^\infty  {1\over m} t^{m} Z_{\text{CFT}}(m\tau,m\overline{\tau})\right]
\end{equation}
When the seed partition function has the following form,
\begin{equation}
Z_{\text{CFT}}(\tau,\overline{\tau})=\sum_{l,\overline{l}}\rho(l,\overline{l}) q^{l}\overline{q}^{\overline{l}}
\end{equation}
the generating function of the untwisted partition function can be written as
\begin{equation}
Z_{\text{untwisted}}=\prod_{l,\overline{l}} {1\over \left(1-t q^l \overline{q}^{\overline{l}}\right)^{\rho(l,\overline{l})}}
\end{equation}
which agrees with~\cite{Gaberdiel:2014cha,Belin:2014fna,Gaberdiel:2015mra}. 

It is also easy to calculate a generating function for the partition function of the $2$-cycle twist sector. In the $2$-cycle twist sector, a representation $\mathcal{R}_{\mathcal{P}}=(\mathcal{R}_1,\cdots,\mathcal{R}_k)$ for a primary $\mathcal{P}$ takes the following form. For fixed $m\in\{1,2,\cdots,k\}$,
\begin{equation}
(\mathcal{R}_{j})_{a,\theta}=\begin{cases}
R_{a,\theta} & \mbox{for}\quad (a,\theta)= (1,0)\\
{\tiny\yng(1)} \delta_{j,m}& \mbox{for}\quad (a,\theta)=(2,0)\\
0 &\mbox{otherwise}
\end{cases}\quad\mbox{or}\quad (\mathcal{R}_{j})_{a,\theta}=\begin{cases}
R_{a,\theta} & \mbox{for}\quad (a,\theta)= (1,0)\\
{\tiny\yng(1)} \delta_{j,m} & \mbox{for}\quad (a,\theta)=(2,1)\\
0 &\mbox{otherwise}
\end{cases}
\end{equation}
which means that only one primary corresponding to the label $m$ is twisted by one $2$-cycle. The 2-cycle twist partition function is obtained by summing absolute square of the characters over such representations
\begin{equation}
Z^N_{2\text{-cycle twist}}=\sum_{\mathcal{P}}{\sum_{\mathcal{R}_{\mathcal{P}}} }'\left|\Gamma_N(\mathcal{P},\mathcal{R}_{\mathcal{P}})\right|^2
\end{equation}
and the corresponding generating function is given by
\begin{eqnarray}
Z_{2\text{-cycle twist}}&=&\sum_{N=0}^\infty t^N Z^N_{2\text{-cycle twist}}\cr
&=&{t^2\over 2} \left(Z_{CFT}\left({\tau\over 2},{\overline{\tau}\over 2}\right)+Z_{CFT}\left({\tau+1\over 2},{\overline{\tau}+1\over 2}\right)\right)Z_{\text{untwisted}}\cr
&=&t^2\sum_{\substack{l,\overline{l}\\l-\overline{l}\;:\;\text{even}}} \rho(l,\overline{l})q^{{l\over 2}} \overline{q}^{{\overline{l}\over 2} }\prod_{l,\overline{l}} {1\over \left(1-t q^l \overline{q}^{\overline{l}}\right)^{\rho(l,\overline{l})}}
\end{eqnarray}
which is consistent with the supersymmetric version of the 2-cycle twisted partition function~\cite{Maldacena:1999bp,Gaberdiel:2014cha}. In the same way, one can calculate a generating function for $k$-cycle twist partition function.
\begin{eqnarray}
Z_{k\text{-cycle twist}}&=&Z_{\text{untwisted}}{t^k\over k} \sum_{j=0}^{k-1}Z_{CFT}\left({\tau+j\over k},{\overline{\tau}+j\over k}\right)\cr
&=&t^k \sum_{\substack{l,\overline{l}\\l-\overline{l}\equiv 0 \;\text{mod}\; k}}\rho(l,\overline{l})q^{{l\over k}} \overline{q}^{{\overline{l}\over k}} \prod_{l,\overline{l}} {1\over \left(1-t q^l \overline{q}^{\overline{l}}\right)^{\rho(l,\overline{l})}}
\end{eqnarray}

So far, we have considered the partition function of the full sector (the holomorphic and the anti-holomorphic parts) so that we have summed the absolute square of the orbifold characters over all possible Young tableaux. We now consider orbifold of chiral sector. As mentioned in Section~\ref{sec:hilbert space}, we cannot have the phase factor. In addition, one has to consider only fully symmetric representations, that is, the Young tableaux which have (at most) one row. %Recall that the Young tableau $R_{a,\theta}$ is the irreducible representation of $S_{l_{a,\theta}}$. If we restrict ourselves to the subset of Young tableaux of one row, the resulting sector will be invariant under $\prod S_{l_{a,\theta}}$. To compare to \cite{Dijkgraaf:1996xw,Gaberdiel:2014cha,Belin:2014fna}, we will consider chiral primaries. 

First of all, the generating function of the untwisted chiral partition function can be written as
\begin{eqnarray}
Z^{\text{chiral}}_{\text{untwisted}}&=&\sum_{N=0}^\infty t^N \sum_{\mathcal{P}}\sum_{\mathcal{R}_j} \Gamma_N(\mathcal{P},\mathcal{R}_{\mathcal{P}})\qquad \mbox{where}\quad (\mathcal{R}_{j})_{a,\theta}=\begin{cases}
{\tiny\yng(2)\cdots \yng(2)} & \mbox{for}\quad (a,\theta)= (1,0)\\
0 &\mbox{for}\quad (a,\theta)\neq(1,0)\\
\end{cases}\cr
&=&\prod_{p} \left[\sum_{N=0}^\infty  t^{N} P_{N}(p;1,0,\overbrace{{\tiny\yng(2)\cdots\yng(2)}}^{N})\right]
\end{eqnarray}
Recall that the Schur polynomial for the irreducible representation $\overbrace{{\tiny\yng(2)\cdots\yng(2)}}^{N}$ is given by
\begin{equation}
P_N(\;{\overbrace{\tiny\yng(2)\cdots\yng(2)}^{N}};\{x_m\})=\sum_{\{\lambda_m\}\vdash N }  \prod_{m=1}^N {x_m^{\lambda_m}\over \lambda_m! m^{\lambda_m}}\label{eq:schur polynomial for symmetric rep}
\end{equation}
Using \eqref{eq:schur polynomial for symmetric rep}, we have
\begin{equation}
Z^{\text{chiral}}_{\text{untwisted}}=\prod_{m=1}^\infty\exp\left[{t^m\over m}\sum_p \chi_p(m\tau)\right]
\end{equation}
If the chiral partition function of the seed CFT takes the form
\begin{equation}
\chi_{\text{CFT}}(\tau)= \sum_p  \chi_p(\tau)=\sum_{l }\rho_l q^{l}
\end{equation}
the generating function of thethe untwisted chiral partition function can be written as
\begin{equation}
Z^{\text{chiral}}_{\text{untwisted}}=\prod_{l} {1\over \left(1-t q^l \right)^{\rho_l}}
\end{equation}
which is the non-supersymmetric version of~\cite{Dijkgraaf:1996xw,Gaberdiel:2014cha,Baggio:2015jxa}. The full chiral partition function of the $S_N$ orbifold can be calculated in the similar way. For this we consider Young tableaux which are zero (the trivial representation) for non-zero $\theta$. {\it i.e.} $R_{a,\theta}=0$ for $\theta\ne 0$ and $R_{a,\theta}={\overbrace{\tiny\yng(1)\cdots\yng(1)}^{n}}$. Recalling the form of the expansion of $\Phi^{(p)}_{a,m;\theta}$ in~\eqref{eq:expansion of phi}, one can easily reproduce the result in~\cite{Dijkgraaf:1996xw}
\begin{equation}
Z^{\text{chiral}}=\prod_{a=1}^\infty \prod_{l} {1\over (1-t^a q^n)^{\rho_{na}}}
\end{equation}

%\subsection{Example : Free Boson}

\acknowledgments

We would like to thank Jean Avan for early work on the permutation orbifold characters. We are grateful to Steven G. Avery for helpful discussions. The work of AJ and JY is supported by the Department of Energy under contract DOE-SC0010010. The work of JY is also supported by the Galkin fellowship at Brown University.

\newpage

\appendix

\section{Examples : Genus One Characters of $S_N$ Orbifolds}\label{app:examples}

\begin{table}[hb!]
\-\hspace{-1em}
{\renewcommand{\arraystretch}{2}
\begin{tabular}{|>{\centering} m{2.3cm}|>{\centering} m{7.5cm}|>{\centering} m{2.5cm} |>{\centering} m{2cm} |}
\hline
$\mathcal{R}$ & $S_2$ orbifold character & $E^{\text{orb}}_2(\mathcal{R})$  & $E^{\text{orb}}_3(\mathcal{R})$ \tabularnewline \hline
$(({\tiny\yng(2)}),(0,0))$  & $P_2(1,0,\;{\tiny\yng(2)}\;)=\frac{1}{2}\left[\left(\chi\left(\tau\right)\right)^2+\chi\left(2\tau\right)\right]$ & $2h_p$ & $\epsilon_{1,0}$  \tabularnewline \hline
$(({\tiny\yng(1,1)}),(0,0))$ & $P_2(1,0,\;{\tiny\yng(1,1)}\;)=\frac{1}{2}\left[\left(\chi\left(\tau\right)\right)^2-\chi\left(2\tau\right)\right]$ & $2h_p$ & $-\epsilon_{1,0}$ \tabularnewline\hline
$((0),({\tiny\yng(1)},0))$ & $P_1(2,0,\;{\tiny\yng(1)}\;)=\frac{1}{2}\left[\chi\left(\frac{\tau}{2}\right)+\chi\left(\frac{\tau+1}{2}\right)\right]$ & $\frac{1}{2}h_p+\frac{1}{16}c$ & $0$ \tabularnewline\hline
$((0),(0,{\tiny\yng(1)}))$  & $P_1(2,1,\;{\tiny\yng(1)}\;)=\frac{1}{2}\left[\chi\left(\frac{\tau}{2}\right)-\chi\left(\frac{\tau+1}{2}\right)\right]$ & $\frac{1}{2}h_p+\frac{1}{16}c+\frac{1}{2}$ & $0$ \tabularnewline \hline
\end{tabular}
}
\caption{Characters for primaries $\langle p,p\rangle$ of $S_2 $ orbifold. }
\end{table}

\begin{table}[hb!]
\-\hspace{-2em}
{\renewcommand{\arraystretch}{2}
\begin{tabular}{|>{\centering} m{2.6cm}|>{\centering} m{9.3cm}|>{\centering} m{2.3cm} |>{\centering} m{1.2cm} |}
\hline
$\mathcal{R}$ & $S_3$ orbifold character & $E^{\text{orb}}_2(\mathcal{R})$ & $E^{\text{orb}}_3(\mathcal{R})$  \tabularnewline \hline
$(({\tiny\yng(3)}),\vec{0}_2,\vec{0}_3)$  & $P_3(1,0,\;{\tiny\yng(3)}\;)=\frac{1}{6}\left[\left(\chi\left(\tau\right)\right)^3+3\chi\left(\tau\right)\chi\left(2\tau\right)+2\chi\left(3\tau\right)\right]$ & $3h_p$ & $3\epsilon_{1,0}$ \tabularnewline \hline
$(({\tiny\yng(2,1)}),\vec{0}_2,\vec{0}_3)$ & $P_3(1,0,\;{\tiny\yng(2,1)}\;)=\frac{1}{3}\left[\left(\chi\left(\tau\right)\right)^3-\chi\left(3\tau\right)\right]$ & $3h_p$ & $0$ \tabularnewline\hline
$(({\tiny\yng(1,1,1)}),\vec{0}_2,\vec{0}_3)$ & $P_3(1,0,\;{\tiny\yng(1,1,1)}\;)=\frac{1}{6}\left[\left(\chi\left(\tau\right)\right)^3-3\chi\left(\tau\right)\chi\left(2\tau\right)-6\chi\left(3\tau\right)\right]$ & $3h_p$ & $-3\epsilon_{1,0}$ \tabularnewline\hline
$(({\tiny\yng(1)}),({\tiny\yng(1)},0),\vec{0}_3)$ & $P_1(1,0,\;{\tiny\yng(1)}\;)P_1(2,0,\;{\tiny\yng(1)}\;)=\chi\left(\tau\right)\frac{1}{2}\left[\chi\left(\frac{\tau}{2}\right)+\chi\left(\frac{\tau+1}{2}\right)\right]$ & $\frac{3}{2}h_p+\frac{1}{16}c$ & $0$ \tabularnewline\hline
$(({\tiny\yng(1)}),(0,{\tiny\yng(1)}),\vec{0}_3)$  & $P_1(1,0,\;{\tiny\yng(1)}\;)P_1(2,1,\;{\tiny\yng(1)}\;)=\chi\left(\tau\right)\frac{1}{2}\left[\chi\left(\frac{\tau}{2}\right)-\chi\left(\frac{\tau+1}{2}\right)\right]$ & $\frac{3}{2}h_p+\frac{1}{16}c+\frac{1}{2}$ & $0$ \tabularnewline \hline
$(\vec{0}_1,\vec{0}_2,({\tiny\yng(1)},0,0))$ & $P_1(3,0,\;{\tiny\yng(1)}\;)=\frac{1}{3}\left[\chi\left(\frac{\tau}{3}\right)+\chi\left(\frac{\tau+1}{3}\right)+\chi\left(\frac{\tau+2}{3}\right)\right]$ & $\frac{1}{3}h_p+\frac{1}{9}c$  & $0$  \tabularnewline\hline
$(\vec{0}_1,\vec{0}_2,(0,{\tiny\yng(1)},0))$ & $P_1(3,1,\;{\tiny\yng(1)}\;)=\frac{1}{3}\left[\chi\left(\frac{\tau}{3}\right)+e^{-\frac{2\pi i}{3}}\chi\left(\frac{\tau+1}{3}\right)+e^{\frac{2\pi i}{3}}\chi\left(\frac{\tau+2}{3}\right)\right]$ & $\frac{1}{3}h_p+\frac{1}{9}c+\frac{1}{3}$  & $0$ \tabularnewline\hline
$(\vec{0}_1,\vec{0}_2,(0,0,{\tiny\yng(1)}))$ & $P_1(3,2,\;{\tiny\yng(1)}\;)=\frac{1}{3}\left[\chi\left(\frac{\tau}{3}\right)+e^{\frac{2\pi i}{3}}\chi\left(\frac{\tau+1}{3}\right)+e^{-\frac{2\pi i}{3}}\chi\left(\frac{\tau+2}{3}\right)\right]$ & $\frac{1}{3}h_p+\frac{1}{9}c+\frac{2}{3}$  & $0$ \tabularnewline\hline
\end{tabular}
}
\caption{Characters for primaries $\langle p,p,p\rangle$ of $S_3 $ orbifold.}
\end{table}

\begin{table}
\-\hspace{-2em}{\renewcommand{\arraystretch}{2}
\begin{tabular}{|>{\centering} m{3.1cm}|>{\centering} m{9.0cm}|>{\centering} m{2.1cm} |>{\centering} m{1.2cm} |}
\hline
$\mathcal{R}$ & $S_4$ orbifold character & $E^{\text{orb}}_2(\mathcal{R})$  & $E^{\text{orb}}_3(\mathcal{R})$  \tabularnewline \hline
$(({\tiny\yng(4)}),\vec{0}_2,\vec{0}_3,\vec{0}_4)$  & $P_4(1,0,\;{\tiny\yng(4)}\;)=\frac{1}{4!}\left[\left(\chi(\tau)\right)^4+6\chi(2\tau)\left(\chi(\tau)\right)^2\right.$ $\left.+8\chi(3\tau)\chi(\tau)+3\left(\chi(2\tau)\right)^2+6\chi(4\tau)\right]$ & $4h_p$ & $6\epsilon_{1,0}$ \tabularnewline \hline
$(({\tiny\yng(3,1)}),\vec{0}_2,\vec{0}_3,\vec{0}_4)$  & $P_4(1,0,\;{\tiny\yng(3,1)}\;)=\frac{1}{4!}\left[3\left(\chi(\tau)\right)^4+6\chi(2\tau)\left(\chi(\tau)\right)^2-3\left(\chi(2\tau)\right)^2-6\chi(4\tau)\right]$ & $4h_p$  & $2\epsilon_{1,0}$  \tabularnewline \hline
$(({\tiny\yng(2,2)}),\vec{0}_2,\vec{0}_3,\vec{0}_4)$  & $P_4(1,0,\;{\tiny\yng(2,2)}\;)=\frac{1}{4!}\left[2\left(\chi(\tau)\right)^4-8\chi(3\tau)\chi(\tau)+6\left(\chi(2\tau)\right)^2\right]$ & $4h_p$  & $0$  \tabularnewline \hline
$(({\tiny\yng(2,1,1)}),\vec{0}_2,\vec{0}_3,\vec{0}_4)$  & $P_4(1,0,\;{\tiny\yng(2,1,1)}\;)=\frac{1}{4!}\left[3\left(\chi(\tau)\right)^4-6\chi(2\tau)\left(\chi(\tau)\right)^2-3\left(\chi(2\tau)\right)^2+6\chi(4\tau)\right]$ & $4h_p$  & $-2\epsilon_{1,0}$  \tabularnewline \hline
$(({\tiny\yng(1,1,1,1)}),\vec{0}_2,\vec{0}_3,\vec{0}_4)$  & $P_4(1,0,\;{\tiny\yng(1,1,1,1)}\;)=\frac{1}{4!}\left[\left(\chi(\tau)\right)^4-6\chi(2\tau)\left(\chi(\tau)\right)^2\right.$ $\left.+8\chi(3\tau)\chi(\tau)+3\left(\chi(2\tau)\right)^2-6\chi(4\tau)\right]$ & $4h_p$  & $-6\epsilon_{1,0}$  \tabularnewline \hline
$(({\tiny\yng(2)}),({\tiny\yng(1)},0),\vec{0}_3,\vec{0}_4)$  & $P_2(1,0,\;{\tiny\yng(2)}\;)P_1(2,0,\;{\tiny\yng(1)}\;)=\frac{1}{2}\left[\left(\chi\left(\tau\right)\right)^2+\chi\left(2\tau\right)\right]\frac{1}{2}\left[\chi\left(\frac{\tau}{2}\right)+\chi\left(\frac{\tau+1}{2}\right)\right]$ & $\frac{5}{2}h_p+\frac{1}{16}c$  & $\epsilon_{1,0}$  \tabularnewline \hline
$(({\tiny\yng(2)}),(0,{\tiny\yng(1)}),\vec{0}_3,\vec{0}_4)$  & $P_2(1,0,\;{\tiny\yng(2)}\;)P_1(2,1,\;{\tiny\yng(1)}\;)=\frac{1}{2}\left[\left(\chi\left(\tau\right)\right)^2+\chi\left(2\tau\right)\right]\frac{1}{2}\left[\chi\left(\frac{\tau}{2}\right)-\chi\left(\frac{\tau+1}{2}\right)\right]$ & $\frac{5}{2}h_p+\frac{1}{16}c+\frac{1}{2}$ & $\epsilon_{1,0}$ \tabularnewline \hline
$(({\tiny\yng(1,1)}),({\tiny\yng(1)},0),\vec{0}_3,\vec{0}_4)$  & $P_2(1,0,\;{\tiny\yng(1,1)}\;)P_1(2,0,\;{\tiny\yng(1)}\;)=\frac{1}{2}\left[\left(\chi\left(\tau\right)\right)^2-\chi\left(2\tau\right)\right]\frac{1}{2}\left[\chi\left(\frac{\tau}{2}\right)+\chi\left(\frac{\tau+1}{2}\right)\right]$ & $\frac{5}{2}h_p+\frac{1}{16}c$  & $-\epsilon_{1,0}$  \tabularnewline \hline
$(({\tiny\yng(1,1)}),(0,{\tiny\yng(1)}),\vec{0}_3,\vec{0}_4)$  & $P_2(1,0,\;{\tiny\yng(1,1)}\;)P_1(2,1,\;{\tiny\yng(1)}\;)=\frac{1}{2}\left[\left(\chi\left(\tau\right)\right)^2-\chi\left(2\tau\right)\right]\frac{1}{2}\left[\chi\left(\frac{\tau}{2}\right)-\chi\left(\frac{\tau+1}{2}\right)\right]$ & $\frac{5}{2}h_p+\frac{1}{16}c+\frac{1}{2}$  & $-\epsilon_{1,0}$  \tabularnewline \hline
\end{tabular}
}
\caption{Characters for primaries $\langle p,p,p,p\rangle$ of $S_4 $ orbifold.}
\end{table}

\begin{table}
\-\hspace{-3.5em}
{\renewcommand{\arraystretch}{2}
\begin{tabular}{|>{\centering} m{3.4cm}|>{\centering} m{9.5cm}|>{\centering} m{2.2cm} |>{\centering} m{1.2cm} |}
\hline
$\mathcal{R}$ & $S_4$ orbifold character & $E^{\text{orb}}_2(\mathcal{R})$ & $E^{\text{orb}}_3(\mathcal{R})$ \tabularnewline \hline
$(({\tiny\yng(1)}),\vec{0}_2,({\tiny\yng(1)},0,0),\vec{0}_4)$  & $P_1(1,0\;{\tiny\yng(1)}\;)P_1(3,0,\;{\tiny\yng(1)}\;)=\chi\left(\tau\right)\frac{1}{3}\left[\chi\left(\frac{\tau}{3}\right)+\chi\left(\frac{\tau+1}{3}\right)+\chi\left(\frac{\tau+2}{3}\right)\right]$ & $\frac{4}{3}h_p+\frac{1}{9}c$ & $0$  \tabularnewline \hline
$(({\tiny\yng(1)}),\vec{0}_2,(0,{\tiny\yng(1)},0),\vec{0}_4)$  & $P_1(1,0\;{\tiny\yng(1)}\;)P_1(3,1,\;{\tiny\yng(1)}\;)=\chi\left(\tau\right)\frac{1}{3}\left[\chi\left(\frac{\tau}{3}\right)+e^{-\frac{2\pi i}{3}}\chi\left(\frac{\tau+1}{3}\right)+e^{\frac{2\pi i}{3}}\chi\left(\frac{\tau+2}{3}\right)\right]$ & $\frac{4}{3}h_p+\frac{1}{9}c+\frac{1}{3}$ & $0$  \tabularnewline \hline
$(({\tiny\yng(1)}),\vec{0}_2,(0,0,{\tiny\yng(1)}),\vec{0}_4)$  & $P_1(1,0\;{\tiny\yng(1)}\;)P_1(3,2,\;{\tiny\yng(1)}\;)=\chi\left(\tau\right)\frac{1}{3}\left[\chi\left(\frac{\tau}{3}\right)+e^{\frac{2\pi i}{3}}\chi\left(\frac{\tau+1}{3}\right)+e^{-\frac{2\pi i}{3}}\chi\left(\frac{\tau+2}{3}\right)\right]$ & $\frac{4}{3}h_p+\frac{1}{9}c+\frac{2}{3}$ & $0$ \tabularnewline \hline
$(\vec{0}_1,({\tiny\yng(2)},0),\vec{0}_3,\vec{0}_4)$  & $P_2(2,0,\;{\tiny\yng(2)}\;)=\frac{1}{2}\left[\left\{\frac{1}{2}\left(\chi\left(\frac{\tau}{2}\right)+\chi\left(\frac{\tau+1}{2}\right)\right)\right\}^2+\frac{1}{2}\left\{\chi\left(\frac{2\tau}{2}\right)+\chi\left(\frac{2\tau+1}{2}\right)\right\}\right]$ & $h_p+\frac{1}{8}c$ & $\epsilon_{2,0}$ \tabularnewline \hline
$(\vec{0}_1,({\tiny\yng(1,1)},0),\vec{0}_3,\vec{0}_4)$  & $P_2(2,0,\;{\tiny\yng(1,1)}\;)=\frac{1}{2}\left[\left\{\frac{1}{2}\left(\chi\left(\frac{\tau}{2}\right)+\chi\left(\frac{\tau+1}{2}\right)\right)\right\}^2-\frac{1}{2}\left\{\chi\left(\frac{2\tau}{2}\right)+\chi\left(\frac{2\tau+1}{2}\right)\right\}\right]$ & $h_p+\frac{1}{8}c$  & $-\epsilon_{2,0}$ \tabularnewline \hline
$(\vec{0}_1,(0,{\tiny\yng(2)}),\vec{0}_3,\vec{0}_4)$  & $P_2(2,1,\;{\tiny\yng(2)}\;)=\frac{1}{2}\left[\left\{\frac{1}{2}\left(\chi\left(\frac{\tau}{2}\right)-\chi\left(\frac{\tau+1}{2}\right)\right)\right\}^2+\frac{1}{2}\left\{\chi\left(\frac{2\tau}{2}\right)-\chi\left(\frac{2\tau+1}{2}\right)\right\}\right]$ & $h_p+\frac{1}{8}c+1$ & $\epsilon_{2,1}$  \tabularnewline \hline
$(\vec{0}_1,(0,{\tiny\yng(1,1)}),\vec{0}_3,\vec{0}_4)$  & $P_2(2,1,\;{\tiny\yng(1,1)}\;)=\frac{1}{2}\left[\left\{\frac{1}{2}\left(\chi\left(\frac{\tau}{2}\right)-\chi\left(\frac{\tau+1}{2}\right)\right)\right\}^2-\frac{1}{2}\left\{\chi\left(\frac{2\tau}{2}\right)-\chi\left(\frac{2\tau+1}{2}\right)\right\}\right]$ & $h_p+\frac{1}{8}c+1$  & $-\epsilon_{2,1}$ \tabularnewline \hline
$(\vec{0}_1,({\tiny\yng(1)},{\tiny\yng(1)}),\vec{0}_3,\vec{0}_4)$  & $P_1(2,0,\;{\tiny\yng(1)}\;)P_1(2,1,\;{\tiny\yng(1)}\;)=\frac{1}{2}\left[\chi\left(\frac{\tau}{2}\right)-\chi\left(\frac{\tau+1}{2}\right)\right]\frac{1}{2}\left[\chi\left(\frac{\tau}{2}\right)+\chi\left(\frac{\tau+1}{2}\right)\right]$ & $h_p+\frac{1}{8}c+\frac{1}{2}$ & $0$ \tabularnewline \hline
$(\vec{0}_1,\vec{0}_2,\vec{0}_3,({\tiny\yng(1)},0,0,0))$  & $P_1(4,0,\;{\tiny\yng(1)}\;)=\frac{1}{4}\left[\chi\left(\frac{\tau}{4}\right)+\chi\left(\frac{\tau+1}{4}\right)+\chi\left(\frac{\tau+2}{4}\right)+\chi\left(\frac{\tau+3}{4}\right)\right]$ & $\frac{1}{4}h_p+\frac{5}{32}c$  & $0$  \tabularnewline \hline
$(\vec{0}_1,\vec{0}_2,\vec{0}_3,(0,{\tiny\yng(1),0,0}))$  & $P_1(4,1,\;{\tiny\yng(1)}\;)=\frac{1}{4}\left[\chi\left(\frac{\tau}{4}\right)-i\chi\left(\frac{\tau+1}{4}\right)-\chi\left(\frac{\tau+2}{4}\right)+i\chi\left(\frac{\tau+3}{4}\right)\right]$ & $\frac{1}{4}h_p+\frac{5}{32}c+\frac{1}{4}$  & $0$ \tabularnewline \hline
$(\vec{0}_1,\vec{0}_2,\vec{0}_3,(0,0,{\tiny\yng(1),0}))$  & $P_1(4,2,\;{\tiny\yng(1)}\;)=\frac{1}{4}\left[\chi\left(\frac{\tau}{4}\right)-\chi\left(\frac{\tau+1}{4}\right)+\chi\left(\frac{\tau+2}{4}\right)-\chi\left(\frac{\tau+3}{4}\right)\right]$ & $\frac{1}{4}h_p+\frac{5}{32}c+\frac{2}{4}$ & $0$  \tabularnewline \hline
$(\vec{0}_1,\vec{0}_2,\vec{0}_3,(0,0,0,{\tiny\yng(1)}))$  & $P_1(4,3,\;{\tiny\yng(1)}\;)=\frac{1}{4}\left[\chi\left(\frac{\tau}{4}\right)+i\chi\left(\frac{\tau+1}{4}\right)-\chi\left(\frac{\tau+2}{4}\right)-i\chi\left(\frac{\tau+3}{4}\right)\right]$ & $\frac{1}{4}h_p+\frac{5}{32}c+\frac{3}{4}$ & $0$  \tabularnewline \hline
\end{tabular}
}
\caption{Characters for primaries $\langle p,p,p,p\rangle$ of $S_4 $ orbifold.}
\end{table}

\begin{table}
\-\hspace{-4em} 
{\renewcommand{\arraystretch}{2}
\begin{tabular}{|>{\centering} m{3.8cm}|>{\centering} m{10.5cm}|>{\centering} m{1.2cm} |>{\centering} m{1.2cm} |}
\hline
$\mathcal{R}$ & $S_5$ orbifold character & $E^{\text{orb}}_2(\mathcal{R})$  & $E^{\text{orb}}_3(\mathcal{R})$  \tabularnewline \hline
$(({\tiny\yng(5)}),\vec{0}_2,\vec{0}_3,\vec{0}_4,\vec{0}_5)$  & $P_5(1,0,\;{\tiny\yng(5)}\;)=\frac{1}{5!}\left[\left(\chi(\tau)\right)^5+10\chi(2\tau)\left(\chi(\tau)\right)^3\right.$ $\left.+20\chi(3\tau)\left(\chi(\tau)\right)^2+30\chi(4\tau)\chi(\tau)+24\chi(5\tau)\right.$ $\left.+15\left(\chi(2\tau)\right)^2\chi(\tau)+20\chi(2\tau)\chi(3\tau)\right]$ & $5h_p$  & $10\epsilon_{1,0}$  \tabularnewline \hline
$(({\tiny\yng(4,1)}),\vec{0}_2,\vec{0}_3,\vec{0}_4,\vec{0}_5)$  & $P_5(1,0,\;{\tiny\yng(4,1)}\;)=\frac{1}{5!}\left[4\left(\chi(\tau)\right)^5+20\chi(2\tau)\left(\chi(\tau)\right)^3\right.$ $\left.+20\chi(3\tau)\left(\chi(\tau)\right)^2-24\chi(5\tau)-20\chi(2\tau)\chi(3\tau)\right]$ & $5h_p$  & $5\epsilon_{1,0}$ \tabularnewline \hline
$(({\tiny\yng(3,2)}),\vec{0}_2,\vec{0}_3,\vec{0}_4,\vec{0}_5)$  & $P_5(1,0,\;{\tiny\yng(3,2)}\;)=\frac{1}{5!}\left[5\left(\chi(\tau)\right)^5+10\chi(2\tau)\left(\chi(\tau)\right)^3-20\chi(3\tau)\left(\chi(\tau)\right)^2\right.$ $\left.-30\chi(4\tau)\chi(\tau)+15\left(\chi(2\tau)\right)^2\chi(\tau)+20\chi(2\tau)\chi(3\tau)\right]$ & $5h_p$  & $2\epsilon_{1,0}$  \tabularnewline \hline
$(({\tiny\yng(3,1,1)}),\vec{0}_2,\vec{0}_3,\vec{0}_4,\vec{0}_5)$  & $P_5(1,0,\;{\tiny\yng(3,1,1)}\;)=\frac{1}{5!}\left[6\left(\chi(\tau)\right)^5+24\chi(5\tau)-30\left(\chi(2\tau)\right)^2\chi(\tau)\right]$ & $5h_p$  & $0$  \tabularnewline \hline
$(({\tiny\yng(2,2,1)}),\vec{0}_2,\vec{0}_3,\vec{0}_4,\vec{0}_5)$  & $P_5(1,0,\;{\tiny\yng(2,2,1)}\;)=\frac{1}{5!}\left[5\left(\chi(\tau)\right)^5-10\chi(2\tau)\left(\chi(\tau)\right)^3-20\chi(3\tau)\left(\chi(\tau)\right)^2\right.$ $\left.+30\chi(4\tau)\chi(\tau) +15\left(\chi(2\tau)\right)^2\chi(\tau)-20\chi(2\tau)\chi(3\tau)\right]$ & $5h_p$  & $-2\epsilon_{1,0}$  \tabularnewline \hline
$(({\tiny\yng(2,1,1,1)}),\vec{0}_2,\vec{0}_3,\vec{0}_4,\vec{0}_5)$  & $P_5(1,0,\;{\tiny\yng(2,1,1,1)}\;)=\frac{1}{5!}\left[4\left(\chi(\tau)\right)^5-20\chi(2\tau)\left(\chi(\tau)\right)^3\right.$ $\left.+20\chi(3\tau)\left(\chi(\tau)\right)^2-24\chi(5\tau)+20\chi(2\tau)\chi(3\tau)\right]$ & $5h_p$  & $-5\epsilon_{1,0}$  \tabularnewline \hline
$(({\tiny\yng(1,1,1,1,1)}),\vec{0}_2,\vec{0}_3,\vec{0}_4,\vec{0}_5)$  & $P_5(1,0,\;{\tiny\yng(1,1,1,1,1)}\;)=\frac{1}{5!}\left[\left(\chi(\tau)\right)^5-10\chi(2\tau)\left(\chi(\tau)\right)^3+20\chi(3\tau)\left(\chi(\tau)\right)^2\right.$ $\left.-30\chi(4\tau)\chi(\tau) +24\chi(5\tau)+15\left(\chi(2\tau)\right)^2\chi(\tau)\right.$ $\left.-20\chi(2\tau)\chi(3\tau)\right]$ & $5h_p$  & $-10\epsilon_{1,0}$  \tabularnewline \hline
\end{tabular}
}
\caption{Characters for primaries $\langle p,p,p,p,p\rangle$ of $S_5 $ orbifold.}
\end{table}

\begin{table}
\-\hspace{-2em}
{\renewcommand{\arraystretch}{2}
\begin{tabular}{|>{\centering} m{4cm}|>{\centering} m{7.5cm}|>{\centering} m{2.5cm} |>{\centering} m{1.2cm} |}
\hline
$\mathcal{R}$ & $S_5$ orbifold character & $E^{\text{orb}}_2(\mathcal{R})$  & $E^{\text{orb}}_3(\mathcal{R})$  \tabularnewline \hline
$(({\tiny\yng(3)}),({\tiny\yng(1)},0),\vec{0}_3,\vec{0}_4,\vec{0}_5)$  & $P_3(1,0,\;{\tiny\yng(3)}\;)P_1(2,0,\;{\tiny\yng(1)}\;)=\frac{1}{6}\left[\left(\chi\left(\tau\right)\right)^3+3\chi\left(\tau\right)\chi\left(2\tau\right)+2\chi\left(3\tau\right)\right] $ $\times \frac{1}{2}\left[\chi\left(\frac{\tau}{2}\right)+\chi\left(\frac{\tau+1}{2}\right)\right]$ & $\frac{7}{2}h_p+\frac{1}{16}c$  & $3\epsilon_{1,0}$ \tabularnewline \hline
$(({\tiny\yng(3)}),(0,{\tiny\yng(1)}),\vec{0}_3,\vec{0}_4,\vec{0}_5)$  & $P_3(1,0,\;{\tiny\yng(3)}\;)P_1(2,1,\;{\tiny\yng(1)}\;)=\frac{1}{6}\left[\left(\chi\left(\tau\right)\right)^3+3\chi\left(\tau\right)\chi\left(2\tau\right)+2\chi\left(3\tau\right)\right]$ $\times \frac{1}{2}\left[\chi\left(\frac{\tau}{2}\right)-\chi\left(\frac{\tau+1}{2}\right)\right]$ & $\frac{7}{2}h_p+\frac{1}{16}c+\frac{1}{2}$ & $3\epsilon_{1,0}$  \tabularnewline \hline
$(({\tiny\yng(2,1)}),({\tiny\yng(1)},0),\vec{0}_3,\vec{0}_4,\vec{0}_5)$  & $P_3(1,0,\;{\tiny\yng(2,1)}\;)P_1(2,0,\;{\tiny\yng(1)}\;)=\frac{1}{3}\left[\left(\chi\left(\tau\right)\right)^3-\chi\left(3\tau\right)\right]\frac{1}{2}\left[\chi\left(\frac{\tau}{2}\right)+\chi\left(\frac{\tau+1}{2}\right)\right]$ & $\frac{7}{2}h_p+\frac{1}{16}c$  & $0$  \tabularnewline \hline
$(({\tiny\yng(2,1)}),(0,{\tiny\yng(1)}),\vec{0}_3,\vec{0}_4,\vec{0}_5)$  & $P_3(1,0,\;{\tiny\yng(2,1)}\;)P_1(2,1,\;{\tiny\yng(1)}\;)=\frac{1}{3}\left[\left(\chi\left(\tau\right)\right)^3-\chi\left(3\tau\right)\right]\frac{1}{2}\left[\chi\left(\frac{\tau}{2}\right)-\chi\left(\frac{\tau+1}{2}\right)\right]$ & $\frac{7}{2}h_p+\frac{1}{16}c+\frac{1}{2}$  & $0$ \tabularnewline \hline
$(({\tiny\yng(1,1,1)}),({\tiny\yng(1)},0),\vec{0}_3,\vec{0}_4,\vec{0}_5)$  & $P_3(1,0,\;{\tiny\yng(1,1,1)}\;)P_1(2,0,\;{\tiny\yng(1)}\;)=\frac{1}{6}\left[\left(\chi\left(\tau\right)\right)^3-3\chi\left(\tau\right)\chi\left(2\tau\right)+2\chi\left(3\tau\right)\right]$ $\times\frac{1}{2}\left[\chi\left(\frac{\tau}{2}\right)+\chi\left(\frac{\tau+1}{2}\right)\right]$ & $\frac{7}{2}h_p+\frac{1}{16}c$  & $-3\epsilon_{1,0}$  \tabularnewline \hline
$(({\tiny\yng(1,1,1)}),(0,{\tiny\yng(1)}),\vec{0}_3,\vec{0}_4,\vec{0}_5)$  & $P_3(1,0,\;{\tiny\yng(1,1,1)}\;)P_1(2,1,\;{\tiny\yng(1)}\;)=\frac{1}{6}\left[\left(\chi\left(\tau\right)\right)^3-3\chi\left(\tau\right)\chi\left(2\tau\right)+2\chi\left(3\tau\right)\right]$ $\times\frac{1}{2}\left[\chi\left(\frac{\tau}{2}\right)-\chi\left(\frac{\tau+1}{2}\right)\right]$ & $\frac{7}{2}h_p+\frac{1}{16}c+\frac{1}{2}$  & $-3\epsilon_{1,0}$  \tabularnewline \hline
\end{tabular}
}
\caption{Characters for primaries $\langle p,p,p,p,p\rangle$ of $S_5 $ orbifold. }
\end{table}

\begin{table}
\-\hspace{-2em}
{\renewcommand{\arraystretch}{2}
\begin{tabular}{|>{\centering} m{4.0cm}|>{\centering} m{8.0cm}|>{\centering} m{2.0cm} |>{\centering} m{1.2cm} |}
\hline
$\mathcal{R}$ & $S_5$ orbifold character & $E^{\text{orb}}_2(\mathcal{R})$ & $E^{\text{orb}}_3(\mathcal{R})$  \tabularnewline \hline
$(({\tiny\yng(2)}),\vec{0}_2,({\tiny\yng(1)},0,0),\vec{0}_4,\vec{0}_5)$  & $P_2(1,0,\;{\tiny\yng(2)}\;)P_1(3,0,\;{\tiny\yng(1)}\;)=\frac{1}{2}\left[\left(\chi\left(\tau\right)\right)^2+\chi\left(2\tau\right)\right]$ $\times\frac{1}{3}\left[\chi\left(\frac{\tau}{3}\right)+\chi\left(\frac{\tau+1}{3}\right)+\chi\left(\frac{\tau+2}{3}\right)\right]$ & $\frac{7}{3}h_p+\frac{1}{9}c$ & $\epsilon_{1,0}$  \tabularnewline \hline
$(({\tiny\yng(2)}),\vec{0}_2,(0,{\tiny\yng(1),0}),\vec{0}_4,\vec{0}_5)$  & $P_2(1,0,\;{\tiny\yng(2)}\;)P_1(3,1,\;{\tiny\yng(1)}\;)=\frac{1}{2}\left[\left(\chi\left(\tau\right)\right)^2+\chi\left(2\tau\right)\right]$ $\times\frac{1}{3}\left[\chi\left(\frac{\tau}{3}\right)+e^{-\frac{2\pi i}{3}}\chi\left(\frac{\tau+1}{3}\right)+e^{\frac{2\pi i}{3}}\chi\left(\frac{\tau+2}{3}\right)\right]$ & $\frac{7}{3}h_p+\frac{1}{9}c+\frac{1}{3}$ & $\epsilon_{1,0}$  \tabularnewline \hline
$(({\tiny\yng(2)}),\vec{0}_2,(0,0,{\tiny\yng(1)}),\vec{0}_4,\vec{0}_5)$  & $P_2(1,0,\;{\tiny\yng(2)}\;)P_1(3,2,\;{\tiny\yng(1)}\;)=\frac{1}{2}\left[\left(\chi\left(\tau\right)\right)^2+\chi\left(2\tau\right)\right]$ $\times\frac{1}{3}\left[\chi\left(\frac{\tau}{3}\right)+e^{\frac{2\pi i}{3}}\chi\left(\frac{\tau+1}{3}\right)+e^{-\frac{2\pi i}{3}}\chi\left(\frac{\tau+2}{3}\right)\right]$ & $\frac{7}{3}h_p+\frac{1}{9}c+\frac{2}{3}$ & $\epsilon_{1,0}$  \tabularnewline \hline
$(({\tiny\yng(1,1)}),\vec{0}_2,({\tiny\yng(1)},0,0),\vec{0}_4,\vec{0}_5)$  & $P_2(1,0,\;{\tiny\yng(1,1)}\;)P_1(3,0,\;{\tiny\yng(1)}\;)=\frac{1}{2}\left[\left(\chi\left(\tau\right)\right)^2-\chi\left(2\tau\right)\right]$ $\times\frac{1}{3}\left[\chi\left(\frac{\tau}{3}\right)+\chi\left(\frac{\tau+1}{3}\right)+\chi\left(\frac{\tau+2}{3}\right)\right]$ & $\frac{7}{3}h_p+\frac{1}{9}c$  & $-\epsilon_{1,0}$  \tabularnewline \hline
$(({\tiny\yng(1,1)}),\vec{0}_2,(0,{\tiny\yng(1),0}),\vec{0}_4,\vec{0}_5)$  & $P_2(1,0,\;{\tiny\yng(1,1)}\;)P_1(3,1,\;{\tiny\yng(1)}\;)=\frac{1}{2}\left[\left(\chi\left(\tau\right)\right)^2-\chi\left(2\tau\right)\right]$ $\times\frac{1}{3}\left[\chi\left(\frac{\tau}{3}\right)+e^{-\frac{2\pi i}{3}}\chi\left(\frac{\tau+1}{3}\right)+e^{\frac{2\pi i}{3}}\chi\left(\frac{\tau+2}{3}\right)\right]$ & $\frac{7}{3}h_p+\frac{1}{9}c+\frac{1}{3}$  & $-\epsilon_{1,0}$  \tabularnewline \hline
$(({\tiny\yng(1,1)}),\vec{0}_2,(0,0,{\tiny\yng(1)}),\vec{0}_4,\vec{0}_5)$  & $P_2(1,0,\;{\tiny\yng(1,1)}\;)P_1(3,2,\;{\tiny\yng(1)}\;)=\frac{1}{2}\left[\left(\chi\left(\tau\right)\right)^2-\chi\left(2\tau\right)\right]$ $\times\frac{1}{3}\left[\chi\left(\frac{\tau}{3}\right)+e^{\frac{2\pi i}{3}}\chi\left(\frac{\tau+1}{3}\right)+e^{-\frac{2\pi i}{3}}\chi\left(\frac{\tau+2}{3}\right)\right]$ & $\frac{7}{3}h_p+\frac{1}{9}c+\frac{2}{3}$ & $-\epsilon_{1,0}$  \tabularnewline \hline
\end{tabular}
}
\caption{Characters for primaries $\langle p,p,p,p,p\rangle$ of $S_5 $ orbifold. }
\end{table}

\begin{table}
\-\hspace{-3em}
{\renewcommand{\arraystretch}{2}
\begin{tabular}{|>{\centering} m{4.2cm}|>{\centering} m{8.3cm}|>{\centering} m{2.1cm} |>{\centering} m{1.2cm} |}
\hline
$\mathcal{R}$ & $S_5$ orbifold character & $E^{\text{orb}}_2(\mathcal{R})$  & $E^{\text{orb}}_3(\mathcal{R})$ \tabularnewline \hline
$(({\tiny\yng(1)}),\vec{0}_2,\vec{0}_3,({\tiny\yng(1)},0,0,0),\vec{0}_5)$  & $P_1(1,0,\;{\tiny\yng(1)}\;)P_1(4,0,\;{\tiny\yng(1)}\;)=\chi\left(\tau\right)\frac{1}{4}\left[\chi\left(\frac{\tau}{4}\right)+\chi\left(\frac{\tau+1}{4}\right)+\chi\left(\frac{\tau+2}{4}\right)+\chi\left(\frac{\tau+3}{4}\right)\right]$ & $\frac{5}{4}h_p+\frac{5}{32}c$ & $0$  \tabularnewline \hline
$(({\tiny\yng(1)}),\vec{0}_2,\vec{0}_3,(0,{\tiny\yng(1),0,0}),\vec{0}_5)$  & $P_1(1,0,\;{\tiny\yng(1)}\;)P_1(4,1,\;{\tiny\yng(1)}\;)=\chi\left(\tau\right)\frac{1}{4}\left[\chi\left(\frac{\tau}{4}\right)-i\chi\left(\frac{\tau+1}{4}\right)-\chi\left(\frac{\tau+2}{4}\right)+i\chi\left(\frac{\tau+3}{4}\right)\right]$ & $\frac{5}{4}h_p+\frac{5}{32}c+\frac{1}{4}$ & $0$  \tabularnewline \hline
$(({\tiny\yng(1)}),\vec{0}_2,\vec{0}_3,(0,0,{\tiny\yng(1),0}),\vec{0}_5)$  & $P_1(1,0,\;{\tiny\yng(1)}\;)P_1(4,2,\;{\tiny\yng(1)}\;)=\chi\left(\tau\right)\frac{1}{4}\left[\chi\left(\frac{\tau}{4}\right)-\chi\left(\frac{\tau+1}{4}\right)+\chi\left(\frac{\tau+2}{4}\right)-\chi\left(\frac{\tau+3}{4}\right)\right]$ & $\frac{5}{4}h_p+\frac{5}{32}c+\frac{2}{4}$ & $0$  \tabularnewline \hline
$(({\tiny\yng(1)}),\vec{0}_2,\vec{0}_3,(0,0,0{\tiny\yng(1)}),\vec{0}_5)$  & $P_1(1,0,\;{\tiny\yng(1)}\;)P_1(4,3,\;{\tiny\yng(1)}\;)=\chi\left(\tau\right)\frac{1}{4}\left[\chi\left(\frac{\tau}{4}\right)+i\chi\left(\frac{\tau+1}{4}\right)-\chi\left(\frac{\tau+2}{4}\right)-i\chi\left(\frac{\tau+3}{4}\right)\right]$ & $\frac{5}{4}h_p+\frac{5}{32}c+\frac{3}{4}$ & $0$ \tabularnewline \hline
\end{tabular}
}
\caption{Characters for primaries $\langle p,p,p,p,p\rangle$ of $S_5 $ orbifold.}
\end{table}

\begin{table}
\-\hspace{-2em}
{\renewcommand{\arraystretch}{2}
\begin{tabular}{|>{\centering} m{4.5cm}|>{\centering} m{7.5cm}|>{\centering} m{2.0cm} |>{\centering} m{1.2cm} |}
\hline
$\mathcal{R}$ & $S_5$ orbifold character & $E^{\text{orb}}_2(\mathcal{R})$  & $E^{\text{orb}}_3(\mathcal{R})$  \tabularnewline \hline
$(\vec{0}_1,\vec{0}_2,\vec{0}_3,\vec{0}_4,({\tiny\yng(1)},0,0,0,0))$  & $P_1(5,0,\;{\tiny\yng(1)}\;)=\frac{1}{5}\left[\chi\left(\frac{\tau}{5}\right)+\chi\left(\frac{\tau+1}{5}\right)+\chi\left(\frac{\tau+2}{5}\right)\right.$ $\left.+\chi\left(\frac{\tau+3}{5}\right)+\chi\left(\frac{\tau+4}{5}\right)\right]$ & $\frac{1}{5}h_p+\frac{1}{5}c$  & $0$  \tabularnewline \hline
$(\vec{0}_1,\vec{0}_2,\vec{0}_3,\vec{0}_4,(0,{\tiny\yng(1)},0,0,0))$  & $P_1(5,1,\;{\tiny\yng(1)}\;)=\frac{1}{5}\left[\chi\left(\frac{\tau}{5}\right)+\rho^4\chi\left(\frac{\tau+1}{5}\right)\right.$ $\left.+\rho^3\chi\left(\frac{\tau+2}{5}\right)+\rho^2\chi\left(\frac{\tau+3}{5}\right)+\rho\chi\left(\frac{\tau+4}{5}\right)\right]$ & $\frac{1}{5}h_p+\frac{1}{5}c+\frac{1}{5}$  & $0$  \tabularnewline \hline
$(\vec{0}_1,\vec{0}_2,\vec{0}_3,\vec{0}_4,(0,0,{\tiny\yng(1)},0,0))$  & $P_1(5,2,\;{\tiny\yng(1)}\;)=\frac{1}{5}\left[\chi\left(\frac{\tau}{5}\right)+\rho^3\chi\left(\frac{\tau+1}{5}\right)\right.$ $\left.+\rho\chi\left(\frac{\tau+2}{5}\right)+\rho^4\chi\left(\frac{\tau+3}{5}\right)+\rho^2\chi\left(\frac{\tau+4}{5}\right)\right]$ & $\frac{1}{5}h_p+\frac{1}{5}c+\frac{2}{5}$  & $0$  \tabularnewline \hline
$(\vec{0}_1,\vec{0}_2,\vec{0}_3,\vec{0}_4,(0,0,{\tiny\yng(1)},0))$  & $P_1(5,3,\;{\tiny\yng(1)}\;)=\frac{1}{5}\left[\chi\left(\frac{\tau}{5}\right)+\rho^2\chi\left(\frac{\tau+1}{5}\right)\right.$ $\left.+\rho^4\chi\left(\frac{\tau+2}{5}\right)+\rho\chi\left(\frac{\tau+3}{5}\right)+\rho^3\chi\left(\frac{\tau+4}{5}\right)\right]$ & $\frac{1}{5}h_p+\frac{1}{5}c+\frac{3}{5}$  & $0$  \tabularnewline \hline
$(\vec{0}_1,\vec{0}_2,\vec{0}_3,\vec{0}_4,(0,0,0,0,{\tiny\yng(1)}))$  & $P_1(5,4,\;{\tiny\yng(1)}\;)=\frac{1}{5}\left[\chi\left(\frac{\tau}{5}\right)+\rho\chi\left(\frac{\tau+1}{5}\right)\right.$ $\left.+\rho^2\chi\left(\frac{\tau+2}{5}\right)+\rho^3\chi\left(\frac{\tau+3}{5}\right)+\rho^4\chi\left(\frac{\tau+4}{5}\right)\right]$ & $\frac{1}{5}h_p+\frac{1}{5}c+\frac{4}{5}$  & $0$ \tabularnewline \hline
\end{tabular}
}
\caption{Characters for primaries $\langle p,p,p,p,p\rangle$ of $S_5 $ orbifold. $\rho\equiv e^{\frac{2\pi i}{5}}$}
\end{table}

\begin{table}
\-\hspace{-3em}
{\renewcommand{\arraystretch}{1.8}
\begin{tabular}{|>{\centering} m{3.3cm}|>{\centering} m{9.6cm}|>{\centering} m{1.9cm} |>{\centering} m{1.2cm} |}
\hline
$\mathcal{R}$ & $S_5$ orbifold character & $E^{\text{orb}}_2(\mathcal{R})$  & $E^{\text{orb}}_3(\mathcal{R})$  \tabularnewline \hline
$({\tiny\yng(1)},({\tiny\yng(2)},0),\vec{0}_3,\vec{0}_4,\vec{0}_5)$  & $P_1(1,0,\;{\tiny\yng(1)}\;)P_2(2,0,\;{\tiny\yng(2)}\;)=\chi\left(\tau\right)\frac{1}{2}\left[\left[\frac{1}{2}\left(\chi\left(\frac{\tau}{2}\right)+\chi\left(\frac{\tau+1}{2}\right)\right)\right]^2+\frac{1}{2}\left(\chi\left(\frac{2\tau}{2}\right)+\chi\left(\frac{2\tau+1}{2}\right)\right)\right]$ & $2h_p+\frac{1}{8}c$  & $\epsilon_{2,0}$  \tabularnewline \hline
$({\tiny\yng(1)},({\tiny\yng(1,1)},0),\vec{0}_3,\vec{0}_4,\vec{0}_5)$  & $P_1(1,0,\;{\tiny\yng(1)}\;)P_2(2,0,\;{\tiny\yng(1,1)}\;)=\chi\left(\tau\right)\frac{1}{2}\left[\left[\frac{1}{2}\left(\chi\left(\frac{\tau}{2}\right)+\chi\left(\frac{\tau+1}{2}\right)\right)\right]^2-\frac{1}{2}\left(\chi\left(\frac{2\tau}{2}\right)+\chi\left(\frac{2\tau+1}{2}\right)\right)\right]$ & $2h_p+\frac{1}{8}c$   & $-\epsilon_{2,0}$  \tabularnewline \hline
$({\tiny\yng(1)},(0,{\tiny\yng(2)}),\vec{0}_3,\vec{0}_4,\vec{0}_5)$  & $P_1(1,0,\;{\tiny\yng(1)}\;)P_2(2,1,\;{\tiny\yng(2)}\;)=\chi\left(\tau\right)\frac{1}{2}\left[\left[\frac{1}{2}\left(\chi\left(\frac{\tau}{2}\right)-\chi\left(\frac{\tau+1}{2}\right)\right)\right]^2+\frac{1}{2}\left(\chi\left(\frac{2\tau}{2}\right)-\chi\left(\frac{2\tau+1}{2}\right)\right)\right]$ & $2h_p+\frac{1}{8}c+1$   & $\epsilon_{2,1}$  \tabularnewline \hline
$({\tiny\yng(1)},(0,{\tiny\yng(1,1)}),\vec{0}_3,\vec{0}_4,\vec{0}_5)$  & $P_1(1,0,\;{\tiny\yng(1)}\;)P_2(2,1,\;{\tiny\yng(1,1)}\;)=\chi\left(\tau\right)\frac{1}{2}\left[\left[\frac{1}{2}\left(\chi\left(\frac{\tau}{2}\right)-\chi\left(\frac{\tau+1}{2}\right)\right)\right]^2-\frac{1}{2}\left(\chi\left(\frac{2\tau}{2}\right)-\chi\left(\frac{2\tau+1}{2}\right)\right)\right]$ & $2h_p+\frac{1}{8}c+1$  & $-\epsilon_{2,1}$  \tabularnewline \hline
$({\tiny\yng(1)},({\tiny\yng(1)},{\tiny\yng(1)}),\vec{0}_3,\vec{0}_4,\vec{0}_5)$  & $P_1(1,0,\;{\tiny\yng(1)}\;)P_1(2,0,\;{\tiny\yng(1)}\;)P_1(2,1,\;{\tiny\yng(1)}\;)=\chi\left(\tau\right)\frac{1}{2}\left[\chi\left(\frac{\tau}{2}\right)-\chi\left(\frac{\tau+1}{2}\right)\right]\frac{1}{2}\left[\chi\left(\frac{\tau}{2}\right)+\chi\left(\frac{\tau+1}{2}\right)\right]$ & $2h_p+\frac{1}{8}c+\frac{1}{2}$   & $0$  \tabularnewline \hline
\end{tabular}
}
\caption{Characters for primaries $\langle p,p,p,p,p\rangle$ of $S_5 $ orbifold.}
\end{table}

\begin{table}
\-\hspace{-3em}
{\renewcommand{\arraystretch}{1.8}
\begin{tabular}{|>{\centering} m{4.3cm}|>{\centering} m{8.2cm}|>{\centering} m{2.3cm} |>{\centering} m{1.2cm} |}
\hline
$\mathcal{R}$ & $S_5$ orbifold character & $E^{\text{orb}}_2(\mathcal{R})$  & $E^{\text{orb}}_3(\mathcal{R})$  \tabularnewline \hline
$(\vec{0}_1,({\tiny\yng(1)},0),({\tiny\yng(1)},0,0),\vec{0}_4,\vec{0}_5)$  & $P_1(2,0,\;{\tiny\yng(1)}\;)P_1(3,0,\;{\tiny\yng(1)}\;)=\frac{1}{2}\left[\chi\left(\frac{\tau}{2}\right)+\chi\left(\frac{\tau+1}{2}\right)\right]$ $\times\frac{1}{3}\left[\chi\left(\frac{\tau}{3}\right)+\chi\left(\frac{\tau+1}{3}\right)+\chi\left(\frac{\tau+2}{3}\right)\right]$ & $\frac{5}{6}h_p+\frac{25}{144}c$  & $0$  \tabularnewline \hline
$(\vec{0}_1,({\tiny\yng(1)},0),(0,{\tiny\yng(1)},0),\vec{0}_4,\vec{0}_5)$  & $P_1(2,0,\;{\tiny\yng(1)}\;)P_1(3,1,\;{\tiny\yng(1)}\;)=\frac{1}{2}\left[\chi\left(\frac{\tau}{2}\right)+\chi\left(\frac{\tau+1}{2}\right)\right]$ $\times\frac{1}{3}\left[\chi\left(\frac{\tau}{3}\right)+e^{-\frac{2\pi i}{3}}\chi\left(\frac{\tau+1}{3}\right)+e^{\frac{2\pi i}{3}}\chi\left(\frac{\tau+2}{3}\right)\right]$ & $\frac{5}{6}h_p+\frac{25}{144}c+\frac{1}{3}$  & $0$  \tabularnewline \hline
$(\vec{0}_1,({\tiny\yng(1)},0),(0,0,{\tiny\yng(1)}),\vec{0}_4,\vec{0}_5)$  & $P_1(2,0,\;{\tiny\yng(1)}\;)P_1(3,2,\;{\tiny\yng(1)}\;)=\frac{1}{2}\left[\chi\left(\frac{\tau}{2}\right)+\chi\left(\frac{\tau+1}{2}\right)\right]$ $\times\frac{1}{3}\left[\chi\left(\frac{\tau}{3}\right)+e^{\frac{2\pi i}{3}}\chi\left(\frac{\tau+1}{3}\right)+e^{-\frac{2\pi i}{3}}\chi\left(\frac{\tau+2}{3}\right)\right]$ & $\frac{5}{6}h_p+\frac{25}{144}c+\frac{2}{3}$  & $0$  \tabularnewline \hline
$(\vec{0}_1,(0,{\tiny\yng(1)}),({\tiny\yng(1)},0,0),\vec{0}_4,\vec{0}_5)$  & $P_1(2,1,\;{\tiny\yng(1)}\;)P_1(3,0,\;{\tiny\yng(1)}\;)=\frac{1}{2}\left[\chi\left(\frac{\tau}{2}\right)-\chi\left(\frac{\tau+1}{2}\right)\right]$ $\times\frac{1}{3}\left[\chi\left(\frac{\tau}{3}\right)+\chi\left(\frac{\tau+1}{3}\right)+\chi\left(\frac{\tau+2}{3}\right)\right]$ & $\frac{5}{6}h_p+\frac{25}{144}c+\frac{1}{2}$  & $0$ \tabularnewline \hline
$(\vec{0}_1,(0,{\tiny\yng(1)}),(0,{\tiny\yng(1)},0),\vec{0}_4,\vec{0}_5)$  & $P_1(2,1,\;{\tiny\yng(1)}\;)P_1(3,1,\;{\tiny\yng(1)}\;)=\frac{1}{2}\left[\chi\left(\frac{\tau}{2}\right)-\chi\left(\frac{\tau+1}{2}\right)\right]$ $\times\frac{1}{3}\left[\chi\left(\frac{\tau}{3}\right)+e^{-\frac{2\pi i}{3}}\chi\left(\frac{\tau+1}{3}\right)+e^{\frac{2\pi i}{3}}\chi\left(\frac{\tau+2}{3}\right)\right]$ & $\frac{5}{6}h_p+\frac{25}{144}c+\frac{5}{6}$  & $0$ \tabularnewline \hline
$(\vec{0}_1,(0,{\tiny\yng(1)}),(0,0,{\tiny\yng(1)}),\vec{0}_4,\vec{0}_5)$  & $P_1(2,1,\;{\tiny\yng(1)}\;)P_1(3,2,\;{\tiny\yng(1)}\;)=\frac{1}{2}\left[\chi\left(\frac{\tau}{2}\right)-\chi\left(\frac{\tau+1}{2}\right)\right]$ $\times\frac{1}{3}\left[\chi\left(\frac{\tau}{3}\right)+e^{\frac{2\pi i}{3}}\chi\left(\frac{\tau+1}{3}\right)+e^{-\frac{2\pi i}{3}}\chi\left(\frac{\tau+2}{3}\right)\right]$ & $\frac{5}{6}h_p+\frac{25}{144}c+\frac{7}{6}$ & $0$ \tabularnewline \hline
\end{tabular}
}
\caption{Characters for primaries $\langle p,p,p,p,p\rangle$ of $S_5 $ orbifold.}
\end{table}

\clearpage


\end{document}